\documentclass[12pt]{iopart}
\expandafter\let\csname equation*\endcsname\relax
\expandafter\let\csname endequation*\endcsname\relax 
\usepackage[utf8]{inputenc}
\usepackage[T1]{fontenc}
\usepackage{mathtools,gensymb,url}         
\usepackage{graphicx}
\usepackage[font=footnotesize,textfont=footnotesize]{subcaption}
\usepackage{acronym}
\usepackage[style=numeric-comp,backend=biber]{biblatex}
\usepackage[normalem]{ulem}     
\usepackage[usenames]{color}
\usepackage{lineno}             
\usepackage{hyperref}
\hypersetup{colorlinks=true,breaklinks=true,urlcolor=blue,
  linkcolor=black,anchorcolor=black,citecolor=black,
  pdfpagemode = UseNone, 
  pdftitle = {Nanofutures paper},   
  pdfsubject = {},
  pdfkeywords = {}
}

\graphicspath{{./}{./img/}{./imglocal/}{./scripts/}}
\newcommand{\revised}[3]{{\color{black}#2}}
\newcommand{\revisednolines}[1]{{\color{black} #1}} 

\usepackage{xcolor}
\definecolor{apricot}{rgb}{0.98, 0.81, 0.69}
\usepackage[color=apricot]{todonotes}

\definecolor{aurometalsaurus}{rgb}{0.43, 0.5, 0.5}

\DeclareSourcemap{
  \maps[datatype=bibtex]{
    \map[overwrite=true]{
      \step[fieldset=url, null]
      \step[fieldset=doi, null]
    }
  }
}

\acrodefplural{RAM}[RAMs]{Random Access Memories}
\acrodefplural{RRAM}[ReRAMs]{Resistive Random Access Memories}
\acrodefplural{STT-MRAM}[STT-MRAMs]{Spin-Transfer Torque Magnetic Random Access Memories}
\acrodef{ADC}[ADC]{Analog to Digital Converter}
\acrodef{ADEX}[AdExp-I\&F]{Adaptive-Exponential Integrate and Fire}
\acrodef{AER}[AER]{Address-Event Representation}
\acrodef{AEX}[AEX]{AER EXtension board}
\acrodef{AE}[AE]{Address-Event}
\acrodef{AFM}[AFM]{Atomic Force Microscope}
\acrodef{AGC}[AGC]{Automatic Gain Control}
\acrodef{AMDA}[AMDA]{AER Motherboard with D/A converters}
\acrodef{ANN}[ANN]{Attractor Neural Network}
\acrodef{API}[API]{Application Programming Interface}
\acrodef{ARM}[ARM]{Advanced RISC Machine}
\acrodef{ASIC}[ASIC]{Application Specific Integrated Circuit}
\acrodef{BCM}[BMC]{Bienenstock-Cooper-Munro}
\acrodef{BD}[BD]{Bundled Data}
\acrodef{BEOL}[BEOL]{Back-end of Line}
\acrodef{BG}[BG]{Bias Generator}
\acrodef{BMI}[BMI]{Brain-Machince Interface}
\acrodef{CAD}[CAD]{Computer Aided Design}
\acrodef{CAM}[CAM]{Content Addressable Memory}
\acrodef{CAVIAR}[CAVIAR]{Convolution AER Vision Architecture for Real-Time}
\acrodef{CFC}[CFC]{Current to Frequency Converter}
\acrodef{CCN}[CCN]{Cooperative and Competitive Network}
\acrodef{CHP}[CHP]{Communicating Hardware Processes}
\acrodef{CNN}[CCN]{Convolutional Neural Network}
\acrodef{CMIM}[CMIM]{Metal-insulator-metal Capacitor}
\acrodef{CMOL}[CMOL]{``Hybrid CMOS nanoelectronic circuits''}
\acrodef{CMOS}[CMOS]{Complementary Metal-Oxide-Semiconductor}
\acrodef{COTS}[COTS]{Commercial Off-The-Shelf}
\acrodef{CPG}[CPG]{Central Pattern Generator}
\acrodef{CPLD}[CPLD]{Complex Programmable Logic Device}
\acrodef{CPU}[CPU]{Central Processing Unit}
\acrodef{CV}[CV]{Coefficient of Variation}
\acrodef{DAC}[DAC]{Digital to Analog Converter}
\acrodef{DAS}[DAS]{Dynamic Auditory Sensor}
\acrodef{DAVIS}[DAVIS]{Dynamic and Active Pixel Vision Sensor}
\acrodef{DBN}[DBN]{Deep Belief Network}
\acrodef{DFA}[DFA]{Deterministic Finite Automaton}
\acrodef{DMA}[DMA]{Direct Memory Access}
\acrodef{DNF}[DNF]{Dynamic Neural Field}
\acrodef{DNN}[DNN]{Deep Neural Network}
\acrodef{DOF}[DOF]{Degrees of Freedom}
\acrodef{DPE}[DPE]{Dynamic Parameter Estimation}
\acrodef{DPI}[DPI]{Differential Pair Integrator}
\acrodef{DRAM}[DRAM]{Dynamic Random Access Memory}
\acrodef{DR}[DR]{Dual Rail}
\acrodef{DSP}[DSP]{Digital Signal Processor}
\acrodef{DVS}[DVS]{Dynamic Vision Sensor}
\acrodef{EBL}[EBL]{Electron Beam Lithography}
\acrodef{EDVAC}[EDVAC]{Electronic Discrete Variable Automatic Computer}
\acrodef{EIN}[EIN]{Excitatory-Inhibitory Network}
\acrodef{EM}[EM]{Expectation Maximization}
\acrodef{EPSC}[EPSC]{Excitatory Post-Synaptic Current}
\acrodef{EPSP}[EPSP]{Excitatory Post-Synaptic Potential}
\acrodef{FDSOI}[FD-SOI]{Fully-Depleted Silicon on Insulator}
\acrodef{FET}[FET]{Field-Effect Transistor}
\acrodef{FFT}[FFT]{Fast Fourier Transform}
\acrodef{FI}[F-I]{Frequency-Current}
\acrodef{FPGA}[FPGA]{Field Programmable Gate Array}
\acrodef{FSA}[FSA]{Finite State Automaton}
\acrodef{FSM}[FSM]{Finite State Machine}
\acrodef{GOPS}[GOPS]{Giga-Operations per Second}
\acrodef{GPU}[GPU]{Graphical Processing Unit}
\acrodef{GUI}[GUI]{Graphical User Interface}
\acrodef{HAL}[HAL]{Hardware Abstraction Layer}
\acrodef{HH}[H\&H]{Hodgkin \& Huxley}
\acrodef{HMM}[HMM]{Hidden Markov Model}
\acrodef{HRS}[HRS]{High-Resistive State}
\acrodef{HR}[HR]{Human Readable}
\acrodef{HSE}[HSE]{Handshaking Expansion}
\acrodef{HW}[HW]{Hardware}
\acrodef{ICT}[ICT]{Information and Communication Technology}
\acrodef{IC}[IC]{Integrated Circuit}
\acrodef{IF2DWTA}[IF2DWTA]{Integrate \& Fire 2--Dimensional WTA}
\acrodef{IFSLWTA}[IFSLWTA]{Integrate \& Fire Stop Learning WTA}
\acrodef{IF}[I\&F]{Integrate-and-Fire}
\acrodef{IMU}[IMU]{Inertial Measurement Unit}
\acrodef{INCF}[INCF]{International Neuroinformatics Coordinating Facility}
\acrodef{INI}[INI]{Institute of Neuroinformatics}
\acrodef{IO}[I/O]{Input/Output}
\acrodef{IoT}[IoT]{Internet of Things}
\acrodef{IPSC}[IPSC]{Inhibitory Post-Synaptic Current}
\acrodef{IPSP}[IPSP]{Inhibitory Post-Synaptic Potential}
\acrodef{IP}[IP]{Intellectual Property}
\acrodef{ISI}[ISI]{Inter-Spike Interval}
\acrodef{JFLAP}[JFLAP]{Java - Formal Languages and Automata Package}
\acrodef{LLC}[LLC]{Low Leakage Cell}
\acrodef{LFP}[LFP]{Local Field Potential}
\acrodef{LNA}[LNA]{Low-Noise Amplifier}
\acrodef{LPF}[LPF]{Low-Pass Filter}
\acrodef{LRS}[LRS]{Low-Resistive State}
\acrodef{LSM}[LSM]{Liquid State Machine}
\acrodef{LTD}[LTD]{Long Term Depression}
\acrodef{LTI}[LTI]{Linear Time-Invariant}
\acrodef{LTP}[LTP]{Long Term Potentiation}
\acrodef{LTU}[LTU]{Linear Threshold Unit}
\acrodef{LUT}[LUT]{Look-Up Table}
\acrodef{MCMC}[MCMC]{Markov-Chain Monte Carlo}
\acrodef{MEMS}[MEMS]{Micro Electro Mechanical System}
\acrodef{MIM}[MIM]{Metal Insulator Metal}
\acrodef{MOSCAP}[MOSCAP]{Metal Oxide Semiconductor Capacitor}
\acrodef{MOSFET}[MOSFET]{Metal Oxide Semiconductor Field-Effect Transistor}
\acrodef{MOS}[MOS]{Metal Oxide Semiconductor}
\acrodef{MRI}[MRI]{Magnetic Resonance Imaging}
\acrodef{NDFSM}[NDFSM]{Non-deterministic Finite State Machine} 
\acrodef{ND}[ND]{Noise-Driven}
\acrodef{NEF}[NEF]{Neural Engineering Framework}
\acrodef{NHML}[NHML]{Neuromorphic Hardware Mark-up Language}
\acrodef{NIL}[NIL]{Nano-Imprint Lithography}
\acrodef{NMDA}[NMDA]{N-Methyl-D-Aspartate}
\acrodef{NME}[NE]{Neuromorphic Engineering}
\acrodef{NSM}[NSM]{Neural State Machine}
\acrodef{OTA}[OTA]{Operational Transconductance Amplifier}
\acrodef{PCB}[PCB]{Printed Circuit Board}
\acrodef{PCM}[PCM]{Phase Change Memory}
\acrodef{PFM}[PFM]{Pulse Frequency Modulation}
\acrodef{PR}[PR]{Production Rule}
\acrodef{PSC}[PSC]{Post-Synaptic Current}
\acrodef{PSTH}[PSTH]{Peri-Stimulus Time Histogram}
\acrodef{QDI}[QDI]{Quasi Delay Insensitive}
\acrodef{RAM}[RAM]{Random Access Memory}
\acrodef{RELU}[ReLu]{Rectified Linear Unit}
\acrodef{RLS}[RLS]{Recursive Least-Squares}
\acrodef{RMSE}[RMSE]{Root Mean Squared-Error}
\acrodef{RMS}[RMS]{Root Mean Squared}
\acrodef{RNN}[RNN]{Recurrent Neural Network}
\acrodef{ROLLS}[ROLLS]{Reconfigurable On-Line Learning Spiking}
\acrodef{RRAM}[ReRAM]{Resistive Random Access Memory}
\acrodef{SAC}[SAC]{Selective Attention Chip}
\acrodef{SCX}[SCX]{Silicon CorteX}
\acrodef{SD}[SD]{Signal-Driven}
\acrodef{SEM}[SEM]{Spike-based Expectation Maximization}
\acrodef{SLAM}[SLAM]{Simultaneous Localization and Mapping}
\acrodef{SOC}[SOC]{System-On-Chip}
\acrodef{SOI}[SOI]{Silicon on Insulator}
\acrodef{SRAM}[SRAM]{Static Random Access Memory}
\acrodef{STDP}[STDP]{Spike-Timing Dependent Plasticity}
\acrodef{STD}[STD]{Short-Term Depression}
\acrodef{STP}[STP]{Short-Term Plasticity}
\acrodef{STT-MRAM}[STT-MRAM]{Spin-Transfer Torque Magnetic Random Access Memory}
\acrodef{STT}[STT]{Spin-Transfer Torque}
\acrodef{SW}[SW]{Software}
\acrodef{TFT}[TFT]{Thin Film Transistor}
\acrodef{USB}[USB]{Universal Serial Bus}
\acrodef{VHDL}[VHDL]{VHSIC Hardware Description Language}
\acrodef{VLSI}[VLSI]{Very Large Scale Integration}
\acrodef{VOR}[VOR]{Vestibulo-Ocular Reflex}
\acrodef{WTA}[WTA]{Winner-Take-All}
\acrodef{XML}[XML]{eXtensible Mark-up Language}
\acrodef{divmod3}[DIVMOD3]{divisibility of a number by 3}
\acrodef{hWTA}[hWTA]{Hard Winner-Take-All}
\acrodef{sWTA}[sWTA]{soft Winner-Take-All}

\begin{document}

\title[DiffMemSyn]{A differential memristive synapse circuit for on-line learning in neuromorphic computing systems}

\author{Manu V Nair, Lorenz K. Muller, and Giacomo Indiveri}

\address{Institute of Neuroinformatics, University of Zurich and ETH Zurich}
\ead{mnair$|$lorenz$|$giacomo@ini.uzh.ch}
\vspace{10pt}
\begin{indented}
\item[]May 2017
\end{indented}

\begin{abstract}
  Spike-based learning with memristive devices in neuromorphic computing architectures typically uses \revised{abstract1}{learning}{\ac{STDP}} circuits that require overlapping pulses from pre- and post-synaptic nodes. This imposes severe constraints on the length of the pulses transmitted in the network, and on the network's throughput. Furthermore, most of these circuits do not decouple the currents flowing through memristive devices from the one stimulating the target neuron. This can be a problem when using devices with high conductance values, because of the resulting large currents. In this paper we propose a novel circuit that decouples the current produced by  the memristive device from the one used to stimulate the post-synaptic neuron, by using a novel differential scheme based on the Gilbert normalizer circuit. We show how this circuit is useful for reducing the effect of variability in the memristive devices, and how it is ideally suited for spike-based learning mechanisms that do not require overlapping pre- and post-synaptic pulses. We demonstrate the features of the proposed synapse circuit with SPICE simulations, and validate its learning properties with high-level behavioral network simulations which use a stochastic gradient descent learning rule in \revised{abstract2}{two classification tasks}{a binary classification}.
\end{abstract}

%
%
%
%
%

\acresetall

\section{Introduction}
\label{sec:intro}

Neuromorphic computing systems typically comprise neuron and synapse circuits arranged in a massively parallel manner to support the emulation of large-scale spiking neural networks~\cite{Chicca_etal14,Park_etal14,Benjamin_etal14,Merolla_etal14a,Indiveri_Liu15,Mitra_etal09,Qiao_etal15,Huayaney_etal16}. In these systems, the bulk of the silicon real-estate is taken up by synaptic circuits, where the memory and computational sites are co-localized~\cite{Indiveri_Liu15}. Consequently, to save area and maximize density, many neuromorphic computing approaches avoid implementing complex synaptic circuits with on-chip learning mechanisms~\cite{Merolla_etal14a,Moradi_etal15,Qiao_Indiveri16}, and resort to training the network on external computers. \revised{intro}{However, these approaches lose}{in addition to losing} the ability to execute on-line ``life-long'' learning \revised{intro2}{and}{, these approaches} require that the network parameters (such as the synaptic weights) be programmed at deployment time. \revised{intro3}{In addition, if}{Furthermore, if} these parameters are stored in \revised{intro4}{}{either on-chip} \ac{SRAM} cells or in \revised{intro5}{}{off-chip} \ac{DRAM} banks, they \revised{intro6}{need to be re-programmed}{are reset whenever power is removed, and have to be uploaded again} \revised{intro7}{every time}{when} the system is \revised{intro8}{reset}{powered-up}. For large  networks~\cite{Merolla_etal14a,Furber_etal14,Schemmel_etal10}, \revised{largenn}{the time required to initialize the system with these parameters can become prohibitive.}{initializing the system with the desired synaptic weights can take a significant amount of time.}

Memristive devices can potentially address these problems by virtue of their compactness and non-volatility~\cite{Ielmini_Waser15}. \revised{intro0MN}{Given their}{Compact resistive memories can be used as binary elements to replace bulky \ac{SRAM} memory cells in digital control and routing circuits; their device physics can be exploited to implement on-line learning mechanisms in the synapse circuits; the non-volatility feature can allow neuromorphic platforms to retain their state between power-cycles. Given these} advantages, several \revised{intro1MN}{neuromorphic}{cross-bar} arrays that use memristive devices \revised{introMN0}{}{in synapses} have been proposed~\revised{refdel}{\cite{Jo_etal10a,Prezioso_etal15,Suri_etal11,Milo_etal16,Pedretti_etal17}.}{\cite{Jo_etal10a,Snider08,Serrano-Gotarredona_etal13,Bi_Poo98,Abbott_Nelson00,Wozniak_etal17,Saighi_etal15,Covi_etal16}.} \revised{soa}{Typically, these approaches propose to use memristive devices in dense synaptic arrays for implementing large-scale neural networks. For instance, \cite{Park_etal15, Prezioso_etal15} describe use of 1R arrays to implement perceptrons  trained by supervised learning protocols such as~\cite{Schiffmann_etal94}. Similarly, in~\cite{Yao_etal17}, the authors train a 1T-1R array to implement perceptrons classifying face images from the Yale face database~\cite{Georghiades_etal97}. In~\cite{Deng_etal15}, the authors use the \ac{RLS} algorithm for training synaptic weights to perform complex tasks such as human motor control. In works such as~\cite{Prezioso_etal15,Deng_etal15}, the authors propose the use of two devices per synaptic element to implement positive and negative weight terms. Other approaches describe synaptic arrays with a 1T-1R synapse elements that learn using classical~\cite{Pedretti_etal17,Garbin_etal15,Suri_etal11,Saighi_etal15,Covi_etal16} or stochastic~\cite{Milo_etal16,Serb_etal16,Bill_Legenstein14} \ac{STDP} learning rules.}{}
\revised{overlapping}{In these arrays the currents used to program the memristive devices can be very large, especially for devices in high-conductance states. This imposes severe restrictions on the power budget, capacitor sizes, and other aspects for the design of ultra-low power memristive-neuromorphic circuits. Moreover,}{However,} \revised{introMN4}{the learning protocols employed in most of these architectures}{use learning schemes, where overlapping pulses are applied across the targeted device such that the effective voltage difference exceeds its switching threshold. These mechanisms} couple the length of the pulses used to transmit signals across the layers of the network with the duration of the pulses required to \revised{introMN5}{program the devices~\cite{Pedretti_etal17,Garbin_etal15}}{implement \ac{STDP}}. This requirement imposes severe constraints on the maximum data throughput of the network, because each row or column in the cross-bar array has to wait for the pre- and post-synaptic pulses to finish, before \revised{overlapping2}{a new one can be sent.}{sending a new one. Furthermore, in most of these designs the currents flowing through the memristive devices are directly fed to the integrating neurons. These current can be very large, especially for devices in high-conductance states. This imposes severe restrictions on the power budget, capacitor sizes, and other design aspects of neuromorphic systems when used for applications that require ultra-low power operation.} In this paper we propose a novel synaptic circuit that \revised{density}{}{is not optimized to maximize density, but} addresses \revised{overlapping3}{at the same time both}{} the large current and overlapping pulses problems. 
To overcome the problem of integrating large currents in the post-synaptic neuron, we propose a novel differential-mode sub-threshold memristive synapse circuit that decouples, normalizes, and re-scales the memristive device current from the one supplied to the post-synaptic neuron. \revised{tosection3}{}{The circuit is based on the classic Gilbert-normalizer element~\cite{Gilbert90,Liu_etal02a}, whose output currents, originally designed for bipolar transistors, but functional also for \acp{MOSFET} operated in the sub-threshold domain~\cite{Liu_etal02a}. The synapse circuit stores its weight as the difference between the conductances of two memristive devices, one representing a positive term and the other representing a negative term. Programming the devices is done in a push-pull manner: to increase the synaptic weight, the conductance of the positive term is increased, and that of the negative term is decreased. The complementary operation is achieved by simultaneously decreasing and increasing the conductances of the positive and negative terms, respectively. The output current produced by this circuit can be scaled to very low values (e.g., in the range of pico Amperes). This reduces the total current driven into the post-synaptic neuron, which can then be implemented with very small capacitors and ultra-low power sub-threshold circuits. The differential operation coupled with the normalizing ability of the circuit has two additional advantages. It reduces the effect of memristive device variability and implements both positive (excitatory) and negative (inhibitory) synapse contributions, effectively doubling the ``high-low'' dynamic range of the synaptic weight.}
To overcome the problem of overlapping pulses in cross-bar architectures, we propose \revised{spikebasedlearning}{an event-based scheme that decouples the duration of the input spikes from the read and update  phases of the target synapse, coupled with the use of}{} a novel spike-based synaptic update mechanism. \revised{todiscussion2}{}{that goes beyond plain \ac{STDP}. Upon the arrival of a  pre-synaptic pulse, a current proportional to the state of the target synapse is transmitted to the post-synaptic neurons, and the state of the memristive devices in the synapse are updated by applying a differential voltage pulse across its terminals. The sign of the differential voltage is set by control signals generated by the post-synaptic learning circuits, which continuously evaluate the post-synaptic neuron activity.} \revised{introMN7}{}{A pulse-shaping circuit decouples the duration of the input spikes from the read and update  phases of the target synapse. This frees the communication bus to transmit spikes from sender nodes to the array, increasing the throughput of the network by use of shared or time-multiplexed communication resources, using protocols such as the one based on the commonly used \ac{AER}~\cite{Deiss_etal98,Boahen00,Merolla_etal07,Chicca_etal07a}.}


In recent years, several algorithms employing spike-triggered learning based on post-synaptic neuronal activity, instead of vanilla \ac{STDP} mechanisms, have been proposed in computational neuroscience literature~\cite{Brader_etal07,Graupner_Brunel12,Urbanczik_Senn14}. Several neuromorphic implementations of these mechanisms have also been realized~\cite{Mitra_etal09,Giulioni_etal12,Qiao_etal15,Huayaney_etal16,Mostafa_etal16}. In this paper, we demonstrate how the proposed differential memristive synapse circuit can be incorporated in a neuromorphic system that employs a \revised{intro2MN}{}{novel} learning circuit based on \revised{similartosuch}{such}{similar} ideas. This circuit is inspired by the biologically plausible learning rule presented in~\cite{Urbanczik_Senn14} and gradient-descent based methods applied to memristive devices~\cite{Nair_Dudek15,Soudry_etal15}. We use these learning circuits to implement a randomized unregulated step descent algorithm, which has been shown to be effective for synaptic elements with limited precision~\cite{Muller_etal17}.

In the following Section, we present the network architecture that is compatible with the proposed differential memristive synapse circuit. In Section~\ref{sec:diff-memr-synapse} we describe the techniques used for sensing and changing the memristive device conductances, and present circuit simulation results that quantify its performance figures. In Section~\ref{sec:sims}, we assess the features of neuromorphic architectures that make use of the proposed circuits and validate them with behavioral simulations in a binary classification task. Finally, \revised{indiscussion}{in the Supplementary material we present extensions and variants to the proposed differential synapse memory cell, including their use in dense 1T-1R cross-bar arrays.}{Section~\ref{sec:discussion} contains the discussion and concluding remarks.}

\section{Neuromorphic architectures for memristive synapses}
\label{sec:arch}
\begin{figure}
	\centering
	\includegraphics[width=0.8\textwidth]{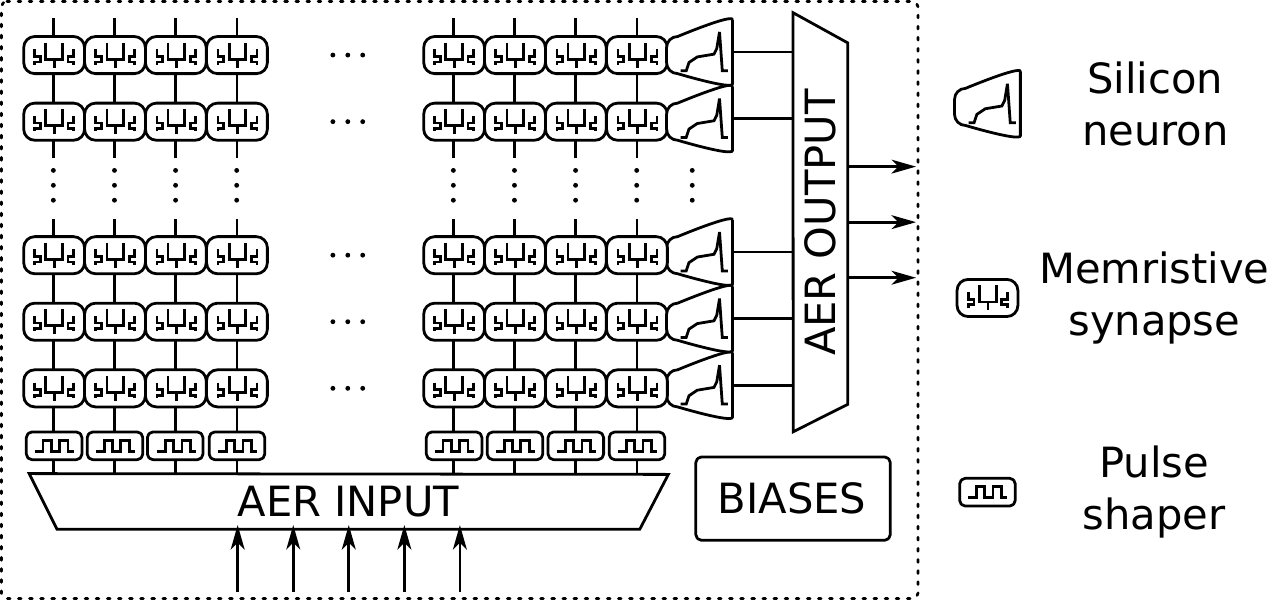}
	\caption{Multi-neuron chip architecture: input \ac{AER} events are converted into read and write pulses sent to multiple memristive synapses; the currents produced by the synaptic circuits are integrated by the neuron assigned for the row; ouput spikes are converted into \ac{AER} events and transmitted off-chip.}
	\label{fig:arch}
\end{figure}

The architecture we propose is composed of an array of synapses and neurons that receive input spikes into columns of synaptic cells, and produce output spikes from the silicon neurons arranged in rows (see Fig.~\ref{fig:arch}). This type of architecture can be integrated within a full-custom neuromorphic \ac{VLSI} chip, or be used as a single-core in multi-core neuromorphic systems~\cite{Moradi_etal15,Merolla_etal14a}. Both the input and output spikes are represented by fast digital pulses that are encoded using the \ac{AER}~\cite{Deiss_etal98,Boahen00,Merolla_etal07,Chicca_etal07a}.

\begin{figure}
	\centering
	\includegraphics[width=0.75\textwidth]{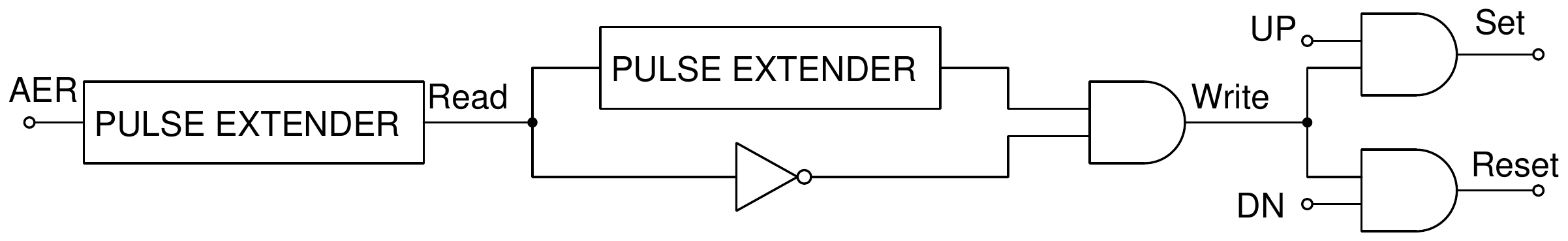}
	\caption{Pulse-shaping circuit for creating a sequence of Read and Write pulses, with each \ac{AER} input event.}
	\label{fig:pulse}
\end{figure}
\revised{input}{On the input side, asynchronous}{Asynchronous} \ac{AER} circuits ensure that events are transmitted as they arrive. \revised{fromintro}{Upon the arrival of a  pre-synaptic address-event a pulse-shaping circuit decouples the duration of the input spikes from the read and update  phases of the target synapse. This frees the communication bus to transmit spikes from sender nodes to the cross-bar array, increasing the throughput of the network by use of shared or time-multiplexed communication resources.}{(on the input side), or are produced (on the output side) in real-time, with minimum latency~\cite{Boahen00}}
The block diagram describing the operation of the pulse-shaping circuits is shown in Fig.~\ref{fig:pulse}. The pulse-shaping circuit \revised{archMN0}{consists}{conists} of two pulse-extender circuits~\cite{Qiao_etal15} and is configured to produce two pulses in quick succession on the arrival of an \ac{AER} event. These pulses sequentially enable the read-mode operation, where the state of the addressed synapse is sensed, \revised{archMN1}{followed by}{and} the write-mode operation, where the state of the memristive devices are increased or decreased, in the targetted synapse. The write-mode operation is directed by the \textsf{UP} and \textsf{DN} control signals produced by the learning circuits \revised{archMN2}{in}{on} the post-synaptic neuron.

\revised{from output}{On the output side a}{A} 1-D arbiter circuit enqueues output events in case of collisions and transmits them on the shared output bus~\cite{Boahen00}.
A programmable bias-generator circuit~\cite{Delbruck_etal10} provides a set of globally-shared temperature-compensated current signals for biasing the analog parameters of the neuromorphic circuits, such as time-constants, refractory periods, or learning rates.
\revised{duplicate}{}{The pulse-shaper circuits arranged on the periphery of the array decouple the length of the incoming \ac{AER} event from the length of the internally-generated pulses required to sense and set conductance values of the memristive devices in the synaptic array. The ability to decouple the \ac{AER} data traffic from the internal pulse lengths allows the system to maximize the spike throughput of the neural network.}




\begin{figure}
	\centering
	\includegraphics[width=\textwidth]{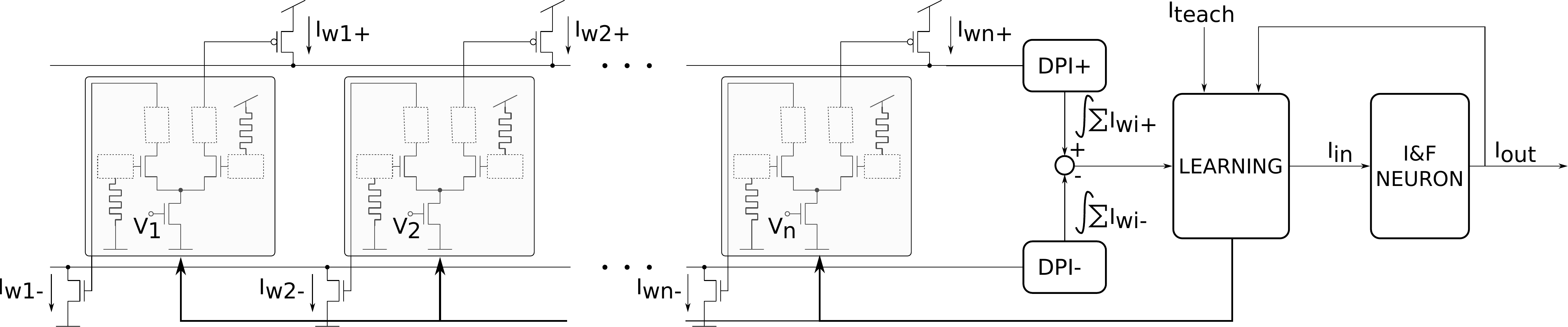}
	\caption{Memristive synapse circuits for on-line learning in a neuromorphic architecture.}
	\label{fig:archnode}
\end{figure}
\revised{parallel}{Address-events target destination columns of the memristive array. By construction, all rows of the stimulated column will process the input event in parallel. Furthermore, the extended read and write pulses typically last longer than the input \ac{AER} event duration. Therefore, a sequential \ac{AER} stimulation of  multiple columns will produce multiple read/write operations across the full array that will overlap in time and operate in parallel.}{}
A block diagram of the circuits present in a single row of the cross-bar architecture illustrated in Fig.~\ref{fig:arch} is shown in Fig.~\ref{fig:archnode}. It comprises multiple synaptic circuits that receive the voltage pulses from the pulse-shaping circuits, two current-mode \ac{DPI} circuits that emulate excitatory and inhibitory synapse dynamics with biologically realistic time constants~\cite{Bartolozzi_Indiveri07a,Bartolozzi_etal06}, a current-mode learning block that implements a spike-driven learning mechanism~\cite{Qiao_etal15,Mitra_etal09}, and an ultra-low-power adaptive \ac{IF} neuron circuit that faithfully reproduces biologically realistic neural dynamics~\cite{Livi_Indiveri09,Indiveri_etal11}. In the read-phase, the synaptic circuit senses the state of the two memristive devices in it, and produces rescaled and normalized differential currents that are driven into the positive and negative \ac{DPI} inputs. The \ac{DPI} circuits integrate these weighted currents and generate a rescaled output current that is driven into a neuron and its learning block. The learning block uses a copy of this ``dendritic'' current to compare it to the net input current, which includes contributions from the neuron and an external source. The external source could represent a teacher signal in supervised learning protocols, or contributions from other neurons in unsupervised learning protocols. Based on this comparison, the learning block evaluates an error signal and produces the \textsf{UP} and \textsf{DN} weight update control signals that are used during the write-mode phase to increase or decrease the weights of the stimulated target synapse. We demonstrate the operation of this architecture with a concrete example in Section~\ref{sec:sims}.

\section{The differential memristive synapse circuit}
\label{sec:diff-memr-synapse}
\revised{fromintro2}{The differential memristive synapse circuit is based on the classic Gilbert-normalizer element~\cite{Gilbert90,Liu_etal02a}, whose output currents, originally designed for bipolar transistors, but functional also for \acp{MOSFET} operated in the sub-threshold domain~\cite{Liu_etal02a}. The synapse circuit stores its weight as the difference between the conductances of two memristive devices, one representing a positive term and the other representing a negative term. Programming the devices is done in a push-pull manner: to increase the synaptic weight, the conductance of the positive term is increased, and that of the negative term is decreased. The complementary operation is achieved by simultaneously decreasing and increasing the conductances of the positive and negative terms, respectively. The output current produced by this circuit, in read-mode, can be scaled to very low values (e.g., in the range of pico Amperes). This reduces the total current driven into the post-synaptic neuron, which can then be implemented with very small capacitors and ultra-low power sub-threshold circuits. The differential operation coupled with the normalizing ability of the circuit has two additional advantages. It reduces the effect of memristive device variability and implements both positive (excitatory) and negative (inhibitory) synapse contributions, effectively doubling the ``high-low'' dynamic range of the synaptic weight.}{} \revised{writecurr}{In  write-mode, the circuit enables programming the memristive devices with programmable current limits, pulse widths, and heights. These can be chosen by the user to optimize the write-mode power consumption depending on the memristive device integrated in the circuit.}{}

\revised{mem_model}{The operating principles of the circuit is independent of the memristive device technology used. It can be used in combination with a wide range of different resistive memory technologies, with arbitrary number of resistive stable states. In this work we assume our \ac{CMOS} circuits can be directly interfaced to $HfO_2$ based devices through post-processing methods, as described in~\cite{Chen_etal17}.}{}

\revised{circuit}{The schematic diagram of the differential memristive synapse circuit is shown in Fig.~\ref{fig:switches}. The circuit is used in a ``read-mode'' phase to measure the conductance of the two memristive devices produce scaled output currents that are conveyed to the downstream current-mode neural processing circuits. It is then operated in a ``write-mode'' for updating the state of the memristive devices via the downstream learning circuit control signals. All $S_i$ \acp{MOSFET} represent switches, with gates controlled by digital signals. Signals with an overline, such as $\overline{X}$, represent the inverted version of the signal $X$. The signal $V_{Read}$ represents the digital voltage used during the read-mode, while the signals $V_{Set}$ and $V_{Reset}$ represent the digital set and reset pulses used in the write-mode to increase/decrease the synaptic weight. The signal $V_b$ is a sub-threshold bias voltage that sets the (sub-threshold) scale of the output currents. The \acp{MOSFET} $Sx$ have dimensions $W/L=5\mu m/0.5 \mu m$; \acp{MOSFET} M1 \& M4 have  $W/L=1\mu m/2 \mu m$, M2 \& M3 have $W/L=0.5\mu m/1 \mu m$, and M5 $W/L=2\mu m/1 \mu m$.}{Figure~\ref{fig:switches} shows the full differential memristive synapse circuit that supports both read- and write-mode circuits}


%
\subsection{Read-mode operation}
\label{sec:read-mode}

\begin{figure}
	\centering
	\includegraphics[width=0.9\textwidth]{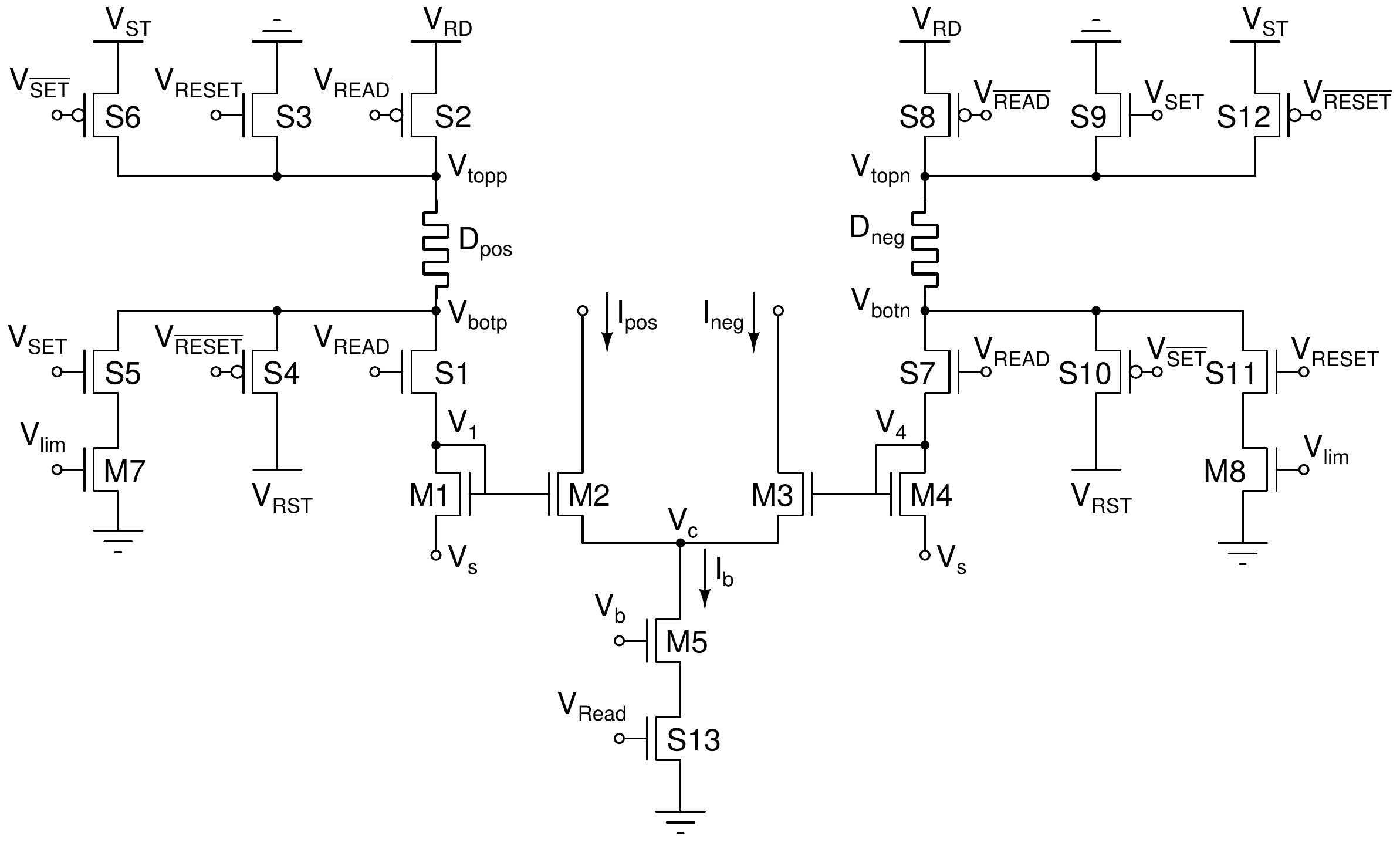}
	\caption{Differential memristive \revisednolines{synaptic} circuit.}
	\label{fig:switches}
\end{figure}

To operate the circuit of Fig.~\ref{fig:switches} in read-mode, the switches S1, S2, S7, and S8 are turned on and all other switches are turned off; the digital control signals $V_{set}$ and $V_{reset}$ are set to logical zero. The current-mode normalizer circuit is implemented by \acp{MOSFET} M1-M6.
In this mode of operation the memristive devices $D_{pos}$ and $D_{neg}$  are connected to corresponding $V_{top}$ and $V_{bot}$  nodes. When the $V_{read}$ pulse is active the currents flowing through the memristive devices are measured and the output currents, $I_{pos}$ and $I_{neg}$, are sent to the excitatory and inhibitory \ac{DPI} circuits, respectively.

Therefore, in this mode of operation, during a $V_{read}$ pulse:
\begin{align}
  I_{Dpos} = & I_{M1}  \; \; \textrm{and} \nonumber\\
  I_{Dneg} = & I_{M4} 
\end{align}
where $I_{Dx}$ is the current through the device $D_x$, and $I_{Mi}$ is the current through the \ac{MOSFET} $Mi$.

For low-power operation, it is desirable to  make $I_{Dx}$ very small. Under this condition, we can assume that the transistors operate in sub-threshold domain. This allows us to analytically derive the relationship between the circuit parameters, and the current flowing through the circuit's output branches.
By writing the sub-threshold equations for a \ac{MOSFET} and equating it to the currents through the resistive devices, we get:
\begin{align}
  (V_{RD} - V_i) = & R_x I_x  \nonumber\\
  I_{Mi}  = &  I_0 e^{\frac{\kappa V_i - V_s}{U_T}}\\
  I_{x}  = & I_{Mi} \nonumber
\end{align}
where $R_x$ represents the resistance of the memristive device $D_x$, $V_{RD}$ the supply voltage provided in ``read-mode'', $V_s$ the source voltage of the input \acp{MOSFET} $M1$ and $M4$, $V_i$ the gate voltage of the \ac{MOSFET} $M_i$, $\kappa$ the sub-threshold slope factor~\cite{Liu_etal02a}, and $U_T$ the thermal voltage. 
By solving for $V_i$:
\begin{equation}
I_x =  I_0 e^{\frac{-\kappa R_xI_x}{U_T}} e^{\frac{\kappa V_{RD} - V_s}{U_T}}
\label{eq:dioderes}
\end{equation}


If \revised{diffmem1MN}{$R_x I_x$}{$I_x$} is sufficiently small, then
\begin{equation}
I_0 e^{\frac{-\kappa R_xI_x}{U_T}} \approx  I_0 (1 - \frac{\kappa}{U_T} R_xI_x)
\end{equation}
so
\begin{equation}
I_x = I_{Mi} = I_0 \frac{1}{e^{-\frac{\kappa V_{RD} -V_s}{U_T}} + \frac{\kappa}{U_T}R_xI_0}
\label{eq:inputcurrent}
\end{equation}

\begin{figure}
  \begin{subfigure}{0.5\textwidth}
    \centering
    \includegraphics[width=\textwidth]{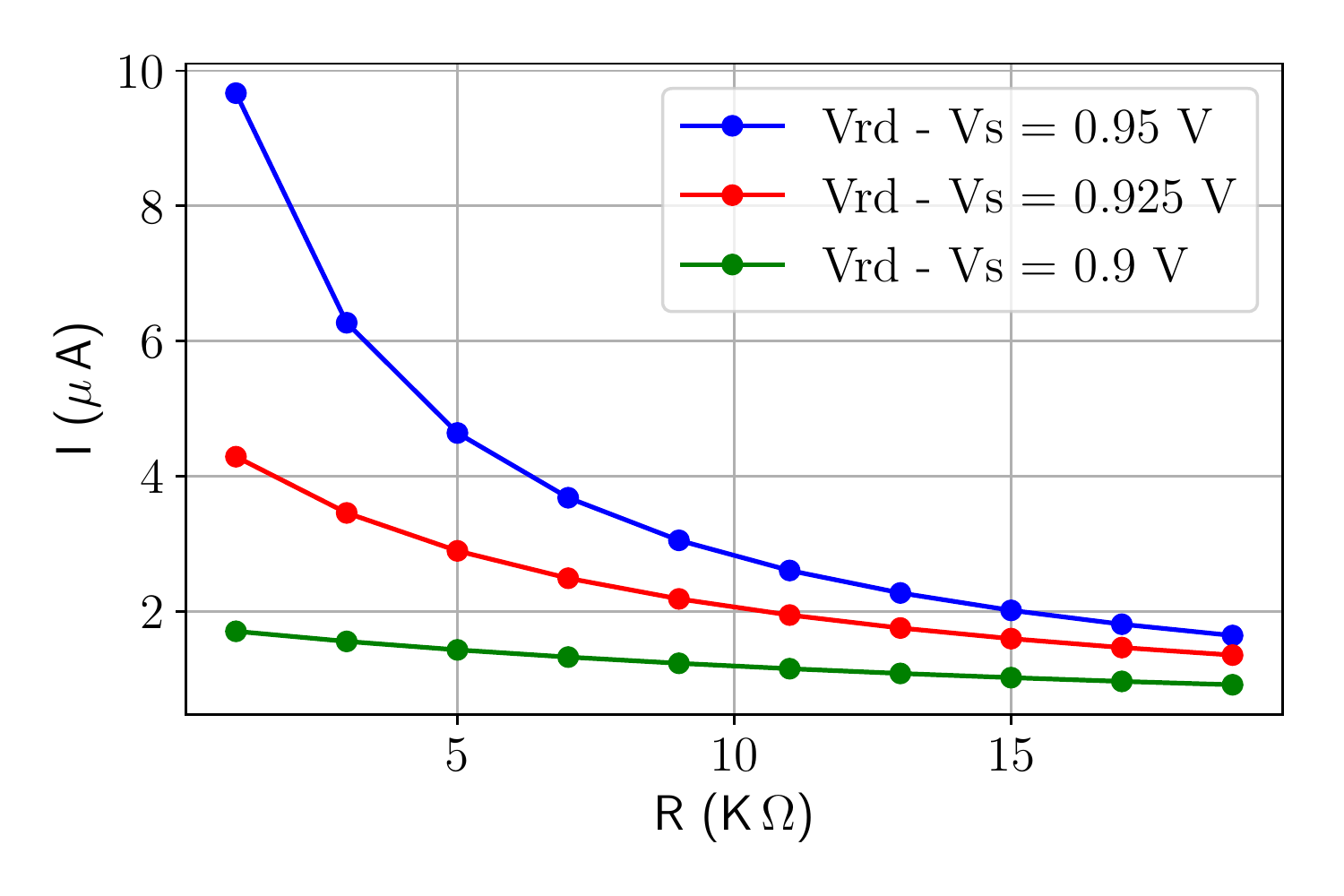}
    \subcaption{}
    \label{fig:analytical-solution}
  \end{subfigure}
  \hfill
  \begin{subfigure}{0.5\textwidth}
    \centering
    \includegraphics[width=\linewidth]{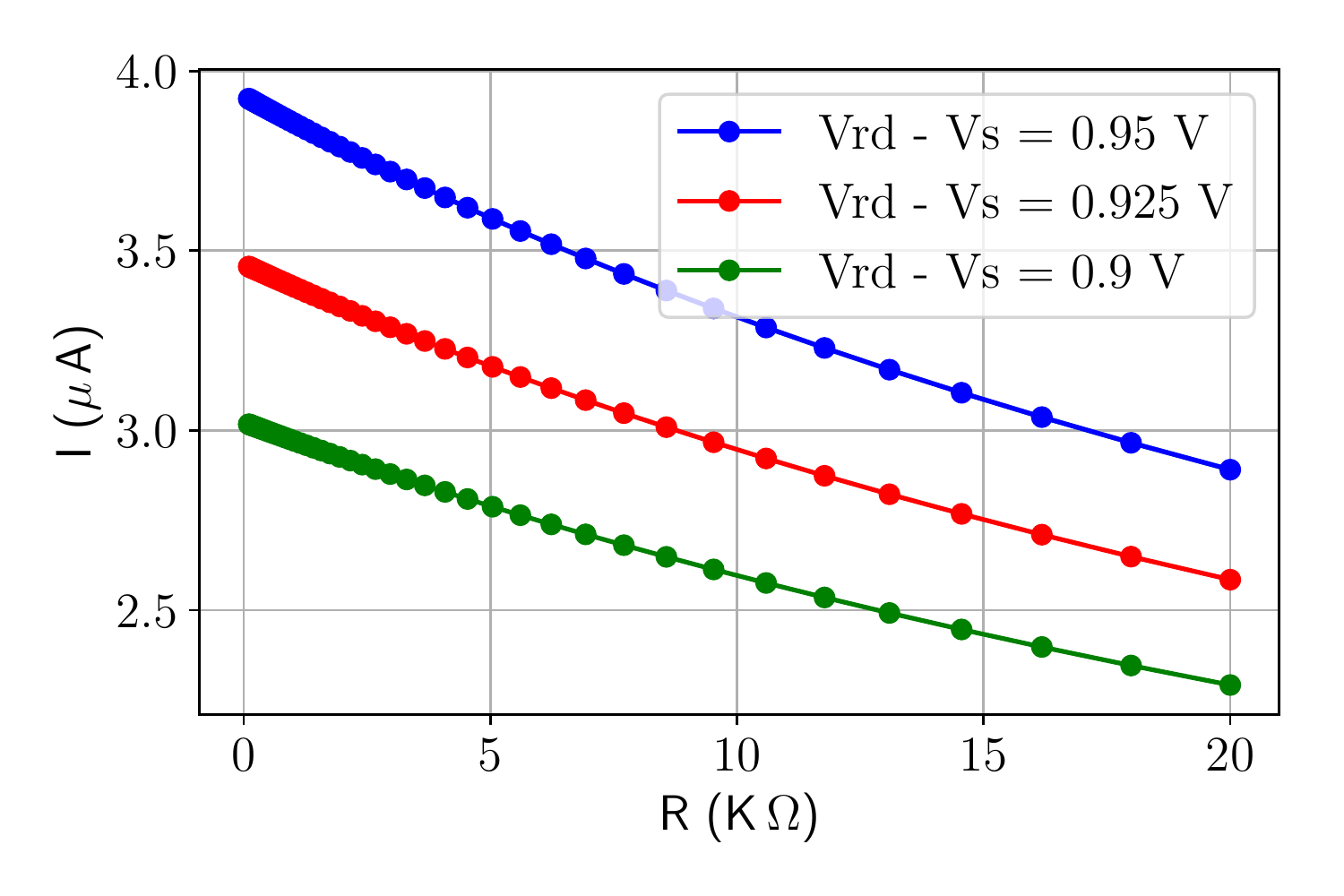}
    \subcaption{}
    \label{fig:diode_sim}
  \end{subfigure}
  \label{fig:inputcurrent}
  \caption{(\subref{fig:analytical-solution}) Theoretical solution of the input current as a function of memristive device resistance, for different values of $V_{RD} - V_s$ settings, with $V_{RD}$ set to 1.8\,V. (\subref{fig:diode_sim}) \revisednolines{SPICE circuit simulations of the same current for a 130\,nm \ac{CMOS} process.}}
\end{figure}

Equation~\ref{eq:inputcurrent} describes how the input current changes with the conductance of the memristive device, and with $V_{RD}$ and $V_s$ voltages. In particular, for large $V_{RD} - V_s$ values, the current is approximately linear with respect to the memristive device conductance, but assumes relatively large values (large values make the circuit less power-efficient). For very small $V_{RD} - V_s$ voltage differences, the circuit produces very small currents that change linearly, but with a very small dependence on the device memristance $R_x$.
The effect of this trade-off is highlighted in Fig.~\ref{fig:analytical-solution}, which plots eq.~\ref{eq:inputcurrent} for different values of $V_s$, with $V_{RD}$ set to 1.8\,V. Figure~\ref{fig:diode_sim} shows circuit simulations results, carried out using a standard 130\,nm \ac{CMOS} process, which support the theoretical analysis.\revised{diffmemMN1}{}{The data shown in Fig.~\ref{fig:diode} was generated using high voltage MOS transistors with M1 and M4 of dimension $W/L = 1\,\mu m/2\,\mu m$.} $V_{RD}$ was set to 1.8\,V, while $V_s$ was swept to obtain the three different $V_{RD} - V_s$ values shown in the figure legend. \revised{fig6mem_model}{In this mode of operation the voltage applied across the memristive device is set low enough to prevent conductance changes. This allows us to model the device as a fixed resistor, and to characterize the circuit as a function of all resistance values between the memristive device's low and high resistance states.}{}

The output currents of the differential memristive circuit are directly proportional to the input currents sensed from the corresponding input branch and scaled by the bias current $I_{b}$. Specifically,if all transistors operate in sub-threshold saturation domain:

\begin{align}
	\label{eq:normalizer}
	I_{M1} & = I_0 e^{\frac{\kappa}{U_T} V_1}e^{ -\frac{V_s}{U_T}}  & I_{pos} & = I_0 e^{\frac{\kappa}{U_T} V_1}e^{ -\frac{V_c}{U_T}}\nonumber\\
	I_{M4} & = I_0 e^{\frac{\kappa}{U_T} V_4}e^{ -\frac{V_s}{U_T}} & I_{neg} & = I_0 e^{\frac{\kappa}{U_T} V_4}e^{ -\frac{V_c}{U_T}}
\end{align}

By solving for $e^{ -\frac{V_c}{U_T}}$ using the extra condition that $I_b = I_{pos} + I_{neg}$,  and replacing terms in eq.~\ref{eq:normalizer}, we obtain:
\begin{align}
	I_{pos} & = I_b\frac{I_{M1}}{I_{M1} + I_{M4}} &   I_{neg} & = I_b\frac{I_{M4}}{I_{M1} + I_{M4}} 
\end{align}
This allows us to produce output currents that are \revised{independent}{scaled versions}{independent} of the currents flowing through the memristive devices, and potentially much smaller, thus enabling the design of ultra low-power current-mode memristive sensing architectures.
In order to ensure proper operation of the differential memristive output normalizing behavior, while minimizing the power dissipated in the input current sensing stage, it is important to have large $V_s$ values, with small $V_{RD} - V_s$ values. 

\begin{figure}
	\centering
	\includegraphics[width=0.65\textwidth]{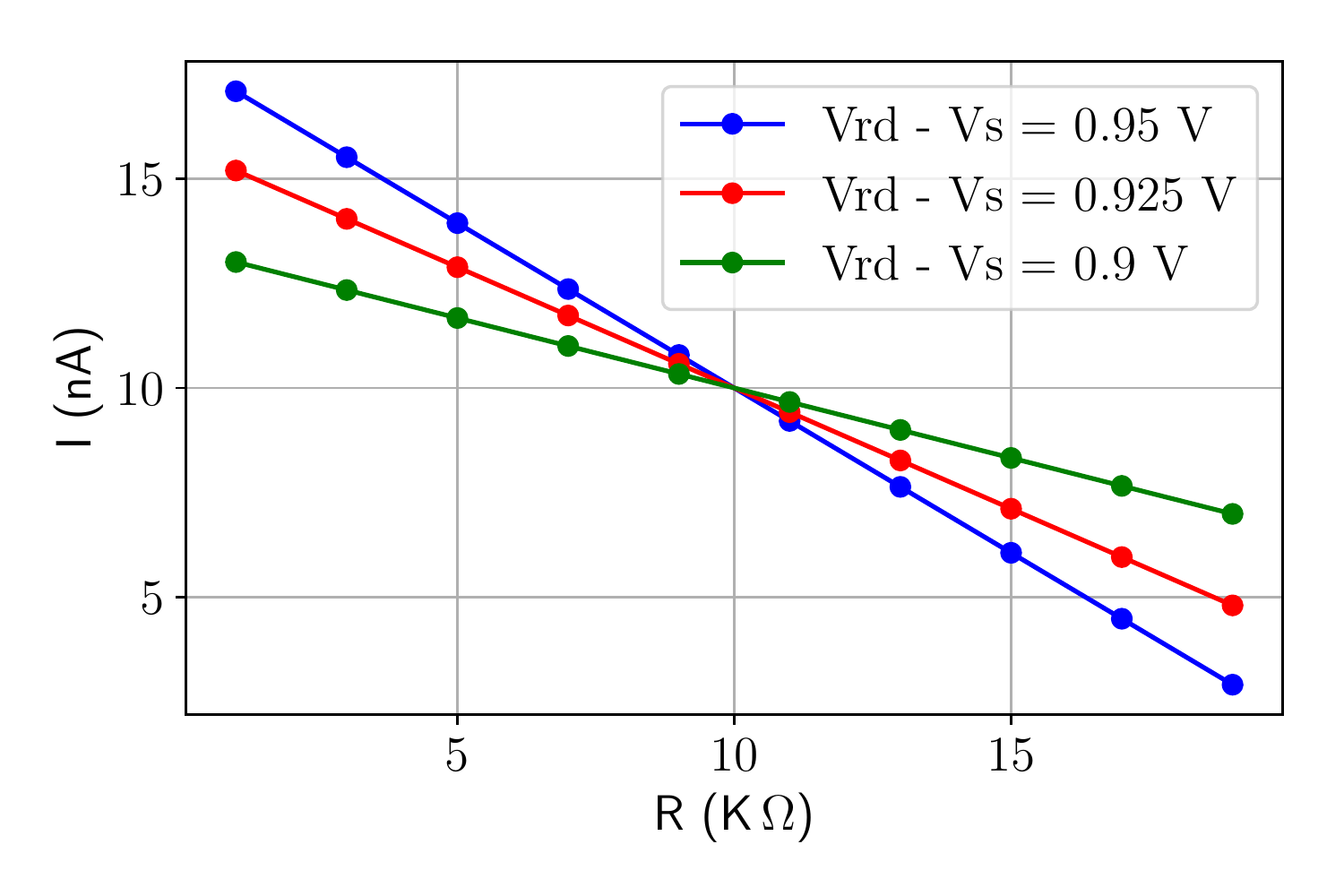}
	\caption{Theoretical normalized output current of the analytic differential memristive circuit as a function of 20 different memristive conductance values, for three different ($V_{RD} - V_s$) settings, with $I_b=20n$\,A, and $V_{rd} = 1.8 V$.}
	\label{fig:normalizer-output}
\end{figure}

\begin{figure}
	\hspace{7em} \includegraphics[width=\textwidth]{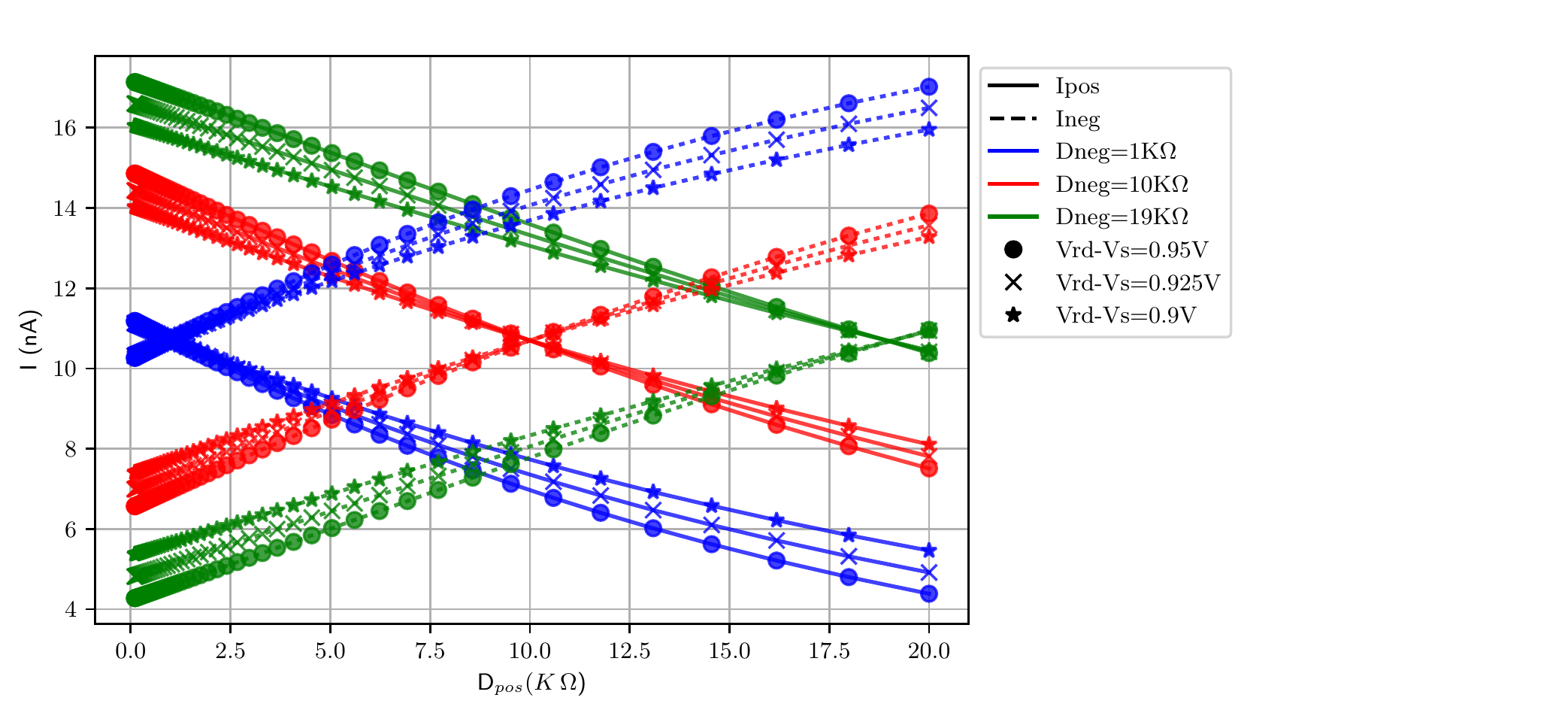}
	\caption{\revisednolines{Circuit simulation results for} the transfer characteristics of the normalizer circuit. \revisednolines{Different colors represent different the circuit response for different values of the $D_{neg}$ resistor. Solid traces represent the current $I_{pos}$ while dashed ones represent $I_{neg}$. Different markers denote different $V_{rd}-V_s$ settings.}}
	\label{fig:rram}
\end{figure}

Figure~\ref{fig:normalizer-output} shows the theoretical normalized output current $I_{pos}$, for a bias current $I_b=20n$\,A, for resistance values of $D_{pos}$ increasing from 1K\,$\Omega$ to 20K\,$\Omega$, and of $D_{neg}$ decreasing proportionally from 20K\,$\Omega$ to 1K\,$\Omega$. A differential current-mode readout circuit that computes $I_{pos}-I_{neg}$ can double the resolution of the conductance/memory state  sensing operation.
More realistic circuit simulation results  \revised{same}{for a}{obtained using the same} 130\,nm \ac{CMOS} process \revised{diffmem2MN}{}{parameters used for the data of Fig.~\ref{fig:diode_sim},} are shown in Fig.~\ref{fig:rram}. To generate this plot, the memristive devices were modelled as resistors and the resistance of $D_{pos}$ was swept from $100 \Omega$ to $20 K\Omega$.
The bias voltage $V_b$ was set to generate a bias current, $I_b$, of 20\,nA. The simulation results show the output of the circuit for different settings of $V_s$ and as a function of different conductance values assumed for the memristive devices. The blue, red, and green traces are the current outputs when the resistance of $D_{neg}$ was set as $1K\Omega$, $10K\Omega$, and $19K\Omega$ respectively. The solid and dashed lines plot $I_{pos}$ and $I_{neg}$ respectively. It can be seen from the plots that the cross over point shifts as the resistance values of $D_{neg}$ change. Note how the linearity of the circuit is improved when $V_{rd} - V_s$ is reduced, at the cost of slightly reduced difference between $I_{pos}$ and $I_{neg}$. 

\subsection{Variability reduction}
\label{sec:mism-norm}
The \revised{complementary}{strategy to use two memristive devices programmed in a complementary fashion and connected to the current-mode normalizer circuit has the additional benefit of significantly reducing the}{current-mode normalizer circuit reduces the} impact of memristive device variability in the output currents.
\begin{figure}
  \begin{subfigure}{0.49\textwidth}
    \centering
    \includegraphics[width=\linewidth]{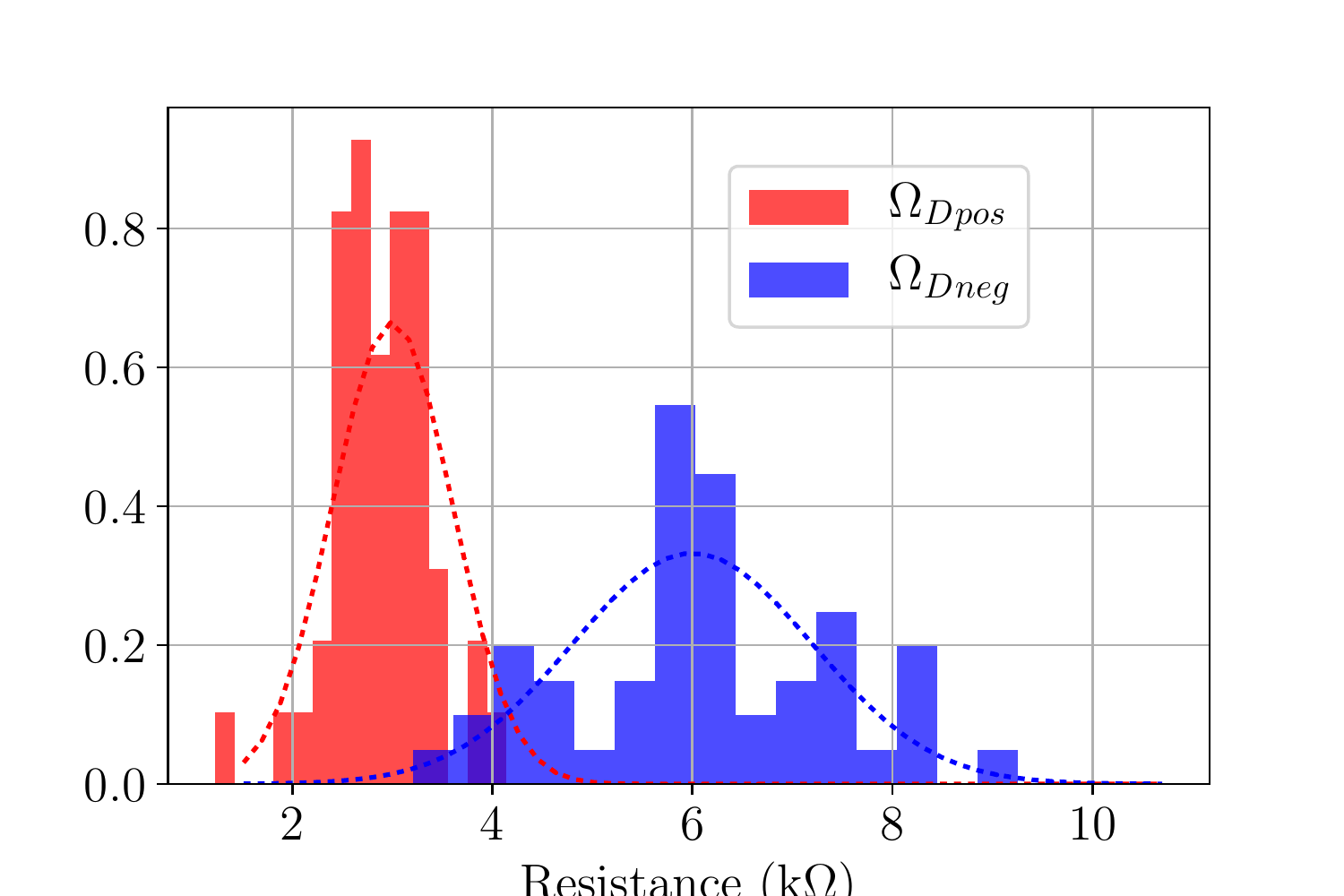}
    \subcaption{}
    \label{fig:mismatch_res_var}
  \end{subfigure}
  \hfill
  \begin{subfigure}{0.49\textwidth}
    \centering
    \includegraphics[width=\linewidth]{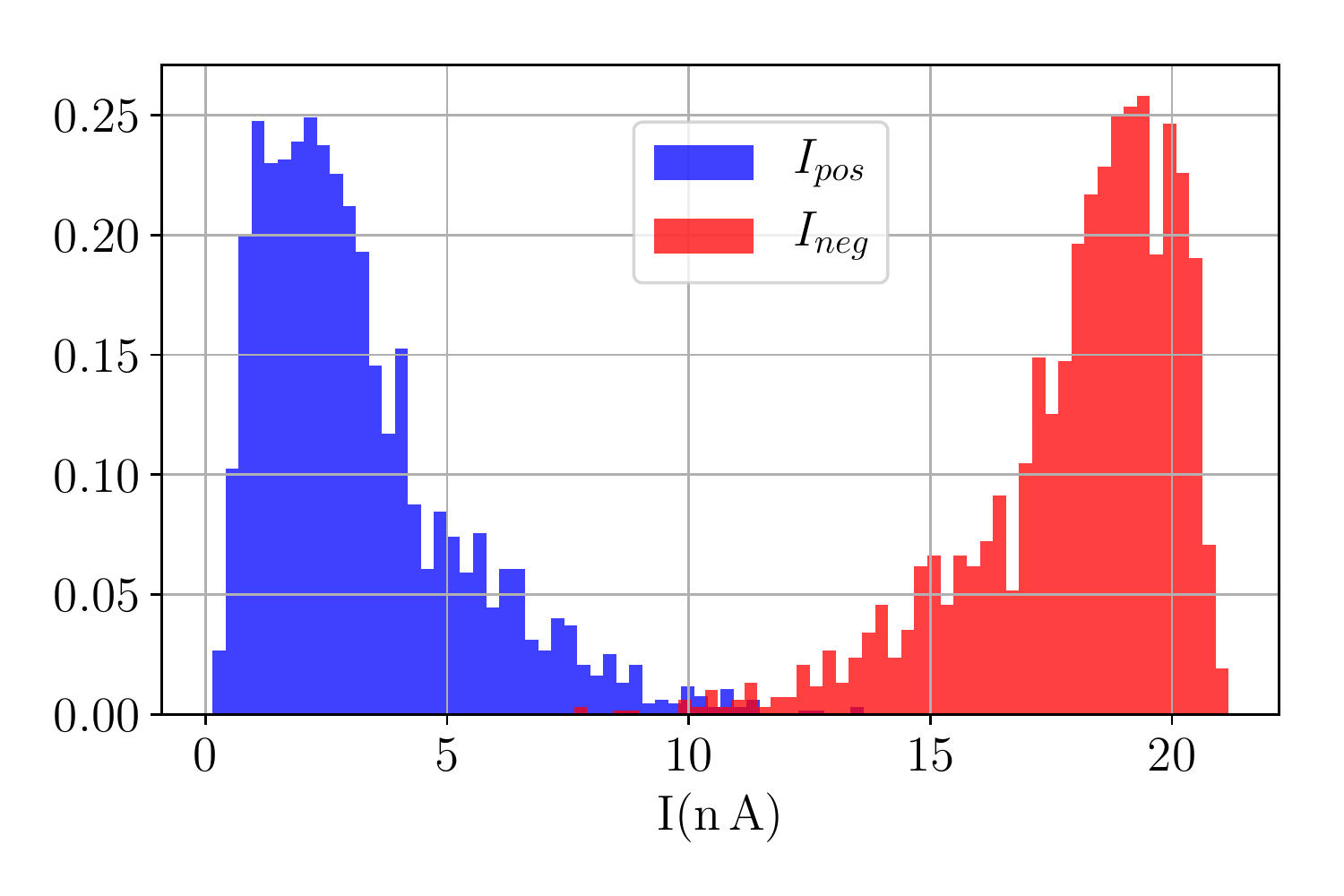}
    \subcaption{}
    \label{fig:mismatch_op_var}
  \end{subfigure}
  \begin{subfigure}{0.49\textwidth}
    \centering
    \includegraphics[width=\textwidth]{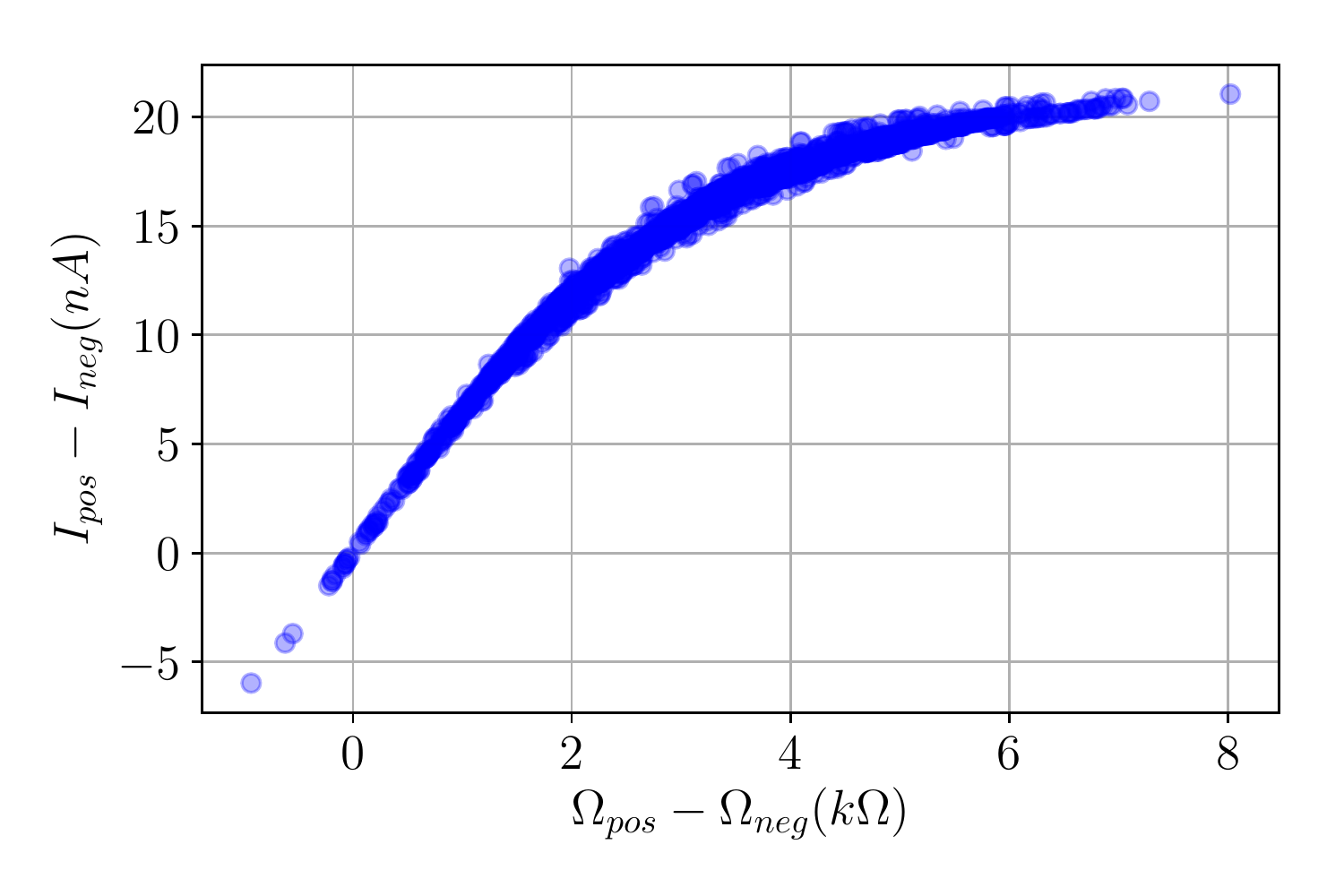}
        \subcaption{}
    \label{fig:tf}
  \end{subfigure}
  \hfill
  \begin{subfigure}{0.49\textwidth}
    \centering
    \includegraphics[width=\textwidth]{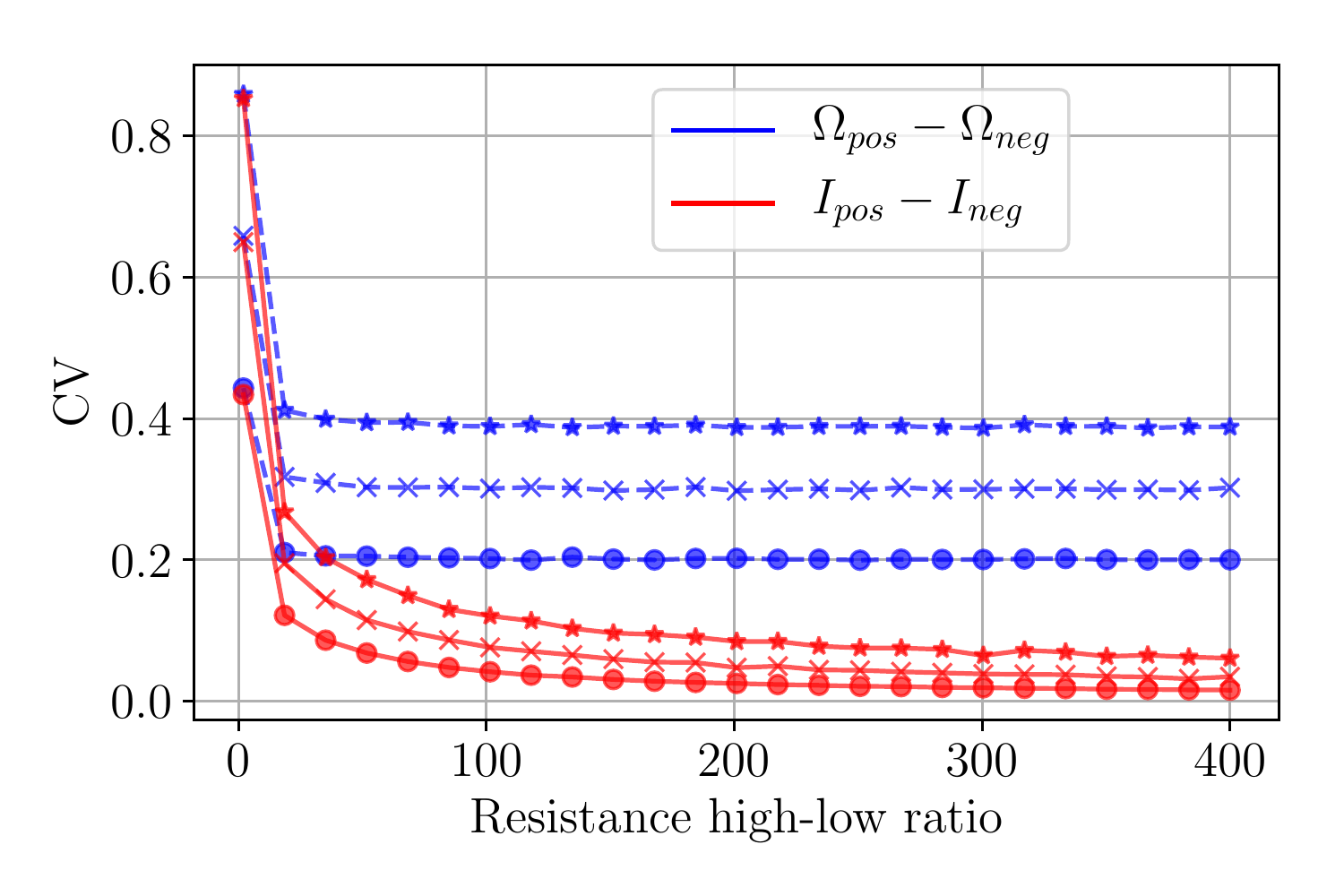}
    \subcaption{}
    \label{fig:tf_cov}
  \end{subfigure}
  \caption{\revisednolines{Effect of memristive device variability on synapse circuits outputs. (\subref{fig:mismatch_res_var}) Distribution of device resistance values (Mean, Std Dev) for $\Omega_{D_{neg}}$ = (2.87$k\Omega$, 490$\Omega$) and  $\Omega_{D_{pos}}$  = (6.12$k\Omega$, 1.3$k\Omega$). The dashed lines represent the sampling distributions. (\subref{fig:mismatch_op_var}) Distribution of output currents for the resistance samples derived from (\subref{fig:mismatch_res_var}).
    (\subref{fig:tf}) Difference of output current samples, versus difference of high-low resistance samples derived from (\subref{fig:mismatch_res_var}).
    (\subref{fig:tf_cov}) Coefficient of variation of output current difference and of resistance difference as a function of different high-low ratios, and for three different values of resistance CV (0.2, 0.3, and 0.4 for the circles, crosses and stars respectively).}}
  \label{fig:mismatch}
\end{figure}
%
\revised{variability-fix}{To demonstrate this effect, we show in Fig.~\ref{fig:mismatch} the results of Monte Carlo simulations in which we compare the variability of the output currents versus the one of the memristive devices.}{In Fig.~\ref{fig:mismatch_res_var}-~\ref{fig:tf}, we show the results of Monte Carlo simulations run on a SPICE simulator that highlight the reduction in variability by comparing the variability in the memristive devices to that of the circuit's output currents.}  In these simulations we set $I_b=20\,nA$ and $V_{S}=0.9\,V$. On the basis of \revised{rram-data}{$HfO_2$}{the} data available from the literature~\cite{Brivio_etal16}, we used conservative figures for the distributions of the memristive device high/low states and their variance. In particular, we sampled resistance values from a Gaussian distribution with (mean,standard deviation) of (6 k$\Omega$,1200 k$\Omega$) and (3 k$\Omega$, 600 $\Omega$) in the high and low resistance states, respectively (see samples in Fig.~\ref{fig:mismatch_res_var}), \revised{output}{and measured the circuit response using such values (see Fig.~\ref{fig:mismatch_op_var}).}{}
 We observed that the histogram of the output currents $I_{pos}$ and $I_{neg}$ are symmetric, illustrating the effect of normalization, with a standard deviation of approximately $2.12\,nA$ for both branches. The normalization circuit effectively compresses the error in output current for large difference between resistances and expands it for small differences as shown in Fig.~\ref{fig:tf}. \revised{cov1}{}{Figure~\ref{fig:tf_cov} illustrates the effect of our circuit on the distribution of the difference between output currents. The asymmetric shape of the distribution arises because the maximum output currents are limited by the tail current of the normalizer.} \revised{conservative}{Even for these conservative values of resistance figures, with a very small high-low ratio, the reduction in the Coefficient of Variation (CV) went from 0.429 for the $\Omega_{pos} - \Omega_{neg}$ to 0.284 for $I_{pos} - I_{neg}$.}{The mean and standard deviation of $I_{pos} - I_{neg}$ and $\Omega_{pos} - \Omega_{neg}$ are $(14.86 nA, 4.22 nA)$ and $(3.24 k\Omega, 1.39 k\Omega)$, respectively. The \ac{CV}, was 0.284 for $I_{pos} - I_{neg}$, and  0.429 for $\Omega_{pos} - \Omega_{neg}$.}
 \revised{var}{For more typical cases, for example with high-low resistance values equivalent to  $100\,K\Omega$ and  $10\,K\Omega$, the same analysis shows a drastic reduction of CV from 0.219 to 0.003.}{}
\revised{CV}{In Fig.~\ref{fig:mismatch} we show a systematic comparison of the CVs between the basic resistance differences and the output current differences, for increasing ratios of high-low states. The comparison was performed running Monte Carlo simulations in which the device high and low resistance states were sampled from a normal distribution with three different coefficients of variation (0.2, 0.3, and 0.4), and the output currents were calculated using the circuit's transfer function derived analytically in Section~\ref{sec:diff-memr-synapse}.}{}

\subsection{Write-mode operation}
\label{sec:write-mode}
The write-mode operation takes place immediately after the read-mode phase, as determined by the sequence of $V_{read}$ and $V_{write}$ pulses generated by the pulse-shaper circuit of Fig.~\ref{fig:pulse}.
In this phase, $V_{read}$ is zero, the $V_{write}$ is high. Furthermore, the switches of two memristive devices (S4-S10) are turned on \revised{complementary2}{in a complementary way,}{} such that the resultant voltage across the memristive devices induce opposite changes in their conductance values. For example, to increase the net output current $(I_{pos} - I_{neg})$, the conductance of $D_{pos}$ is increased and that of $D_{neg}$ is decreased. This is done by enabling the switches S5, S6, S9, and S10 by programming the$V_{set}$ signal to logical one, and $V_{reset}$  to logical zero. This connects $V_{topp}$ to $V_{ST}$,  $V_{botp}$ to ground, $V_{botn}$ to $V_{RST}$, and $V_{topn}$ to ground. Similarly, to decrease $(I_{pos} - I_{neg})$, the $V_{reset}$  signal is to set logical one, and $V_{set}$  is set to logical zero. The \acp{MOSFET} M7 and M8 are current-limiting transistors that protect the devices from damage during programming. The signal $V_{lim}$ is a bias voltage chosen that ensure that the memristive devices are not damaged during the forming operation. To minimize power consumption all switching transistors are turned on only during a read or write pulse.

\revised{writestep}{The pulse shaping circuit of Fig~\ref{fig:pulse} can be tuned to increase or decrease the write pulse duration. So by}{By} programming the length of these pulses and by choosing appropriate values for $V_{ST}$ and $V_{RST}$ voltages, it is possible to use this circuit to produce reliable binary, gradual, or stochastic changes in the memristive devices~\cite{Brivio_etal16,Ielmini_Waser15}.  The mode of operation of the memristive devices and the nature of the changes that should be induced in the memristive device conductance depend on the specific learning algorithm implemented in the learning block of Fig.~\ref{fig:archnode}.


\section{Learning simulations}
\label{sec:sims}


\revised{learning}{
In this section we demonstrate examples of spike-based learning simulations using a learning rule that is ideally suited for implementation in neuromorphic architectures that comprise the memristive synapse proposed. In the first subsection we learn a single low-dimensional pattern with varying contrast, in the second subsection we learn many overlapping high-dimensional patterns.

\subsection{Single pattern binary classification}
\label{sec:single-patt-binary}
Here we show simulation results of two neurons trained to classify an input spike-train by adjusting their synaptic weights. 
We study the performance of such a learning system connected by multiple binary synapses
and compare it to that of a hypothetical 32-bit floating-point precision synapse in the same setting. This illustrates what the performance limitation is with an `ideal' synaptic element for  classifying a finite-rate Poisson train with a leaky integrator neuron.

In this task, the neurons $a$ and $b$ are connected via randomly initialized synapses to two neural populations $p_1$ and $p_2$, which fire with two different average firing rates, with Poissonian statistics. The goal is for neuron $a$ to learn to fire more than neuron $b$, whenever input units from population $p_1$ fire at a higher rate than input units in $p_2$.
To achieve this, we use a supervised training protocol: given the input, we provide a teacher signal to the neuron that should fire more. The teacher signal is represented by a Poissonian spike train sent to the target neurons via an additional, separate, channel.  The spike-based learning algorithm is a discretized version of the one presented in~\cite{Urbanczik_Senn14}. It performs a gradient descent procedure on the difference of the observed and desired neuron firing rates, and it can be readily implemented in mixed signal \ac{CMOS} neuromorphic hardware~\cite{Muller_etal17,Qiao_etal15}.
A detailed description of this learning rule and the parameter values used are provided in the Supplementary Material.

\begin{figure}
 	\centering
 	\includegraphics[width=\textwidth]{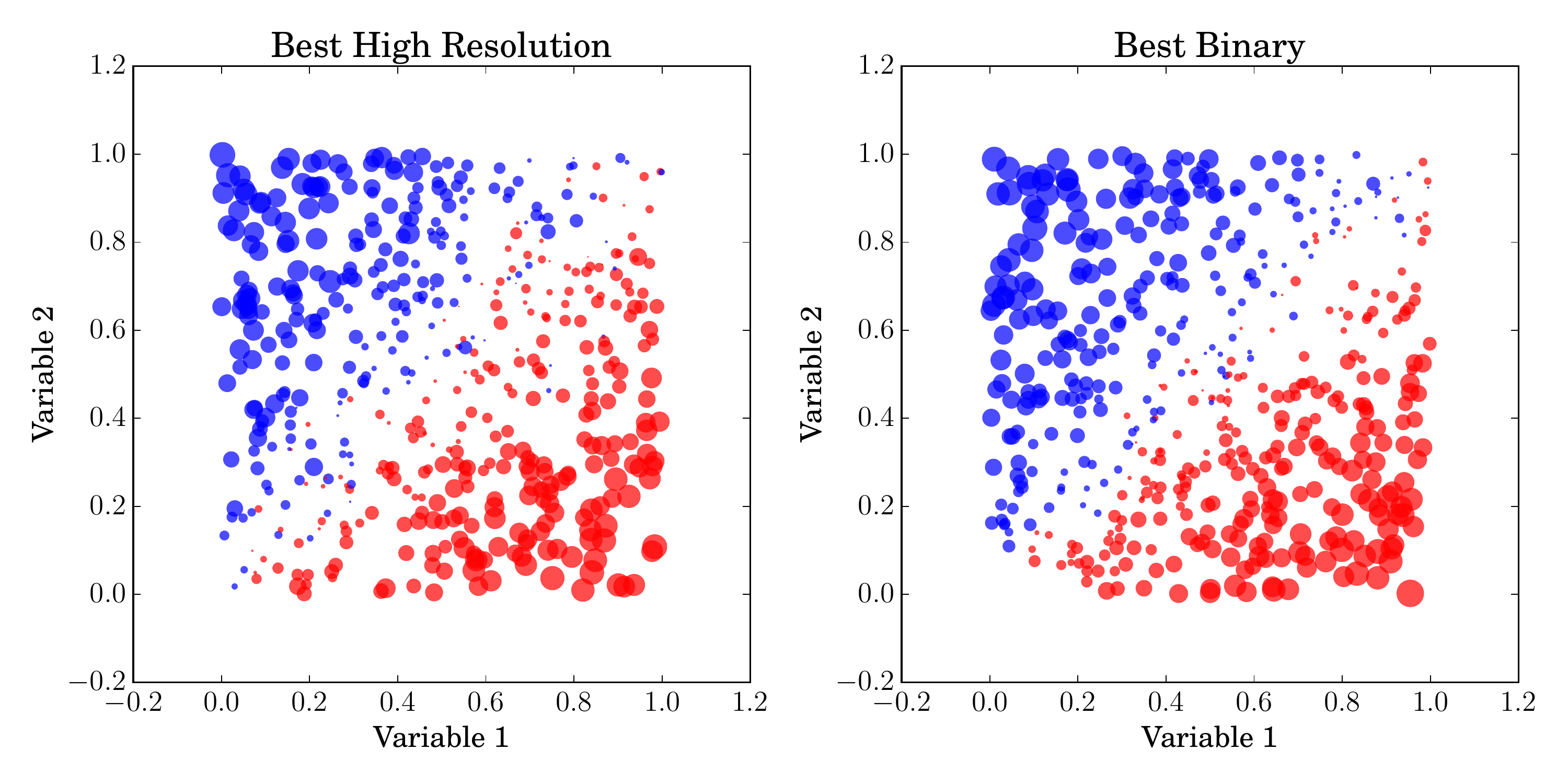}
 	\caption{Comparison of the classification output of the two neuron system with 40 binary weight elements (left) and 40 floating-point resolution synapses (right). The size of the circles indicates the difference in output rates of the two neurons. }
 	\label{fig:LearningClass}
\end{figure}

In Fig.~\ref{fig:LearningClass} we compare the classifications results produced by the best system using a 32-bit floating-point precision synapse with the results obtained simulating the proposed binary synapse in the same setting. Each of the two variables is proportional to the firing rate of the respective subpopulation. When they are equal the pattern has no `constrast', when they are very different it is strongly contrasted. The colors indicate which of the two learning neurons fired more strongly; the ideal solution is a separation of red and blue at a $45\degree$ angle. 
The proposed learning rule finds a good solution to the classification problem, in that the misclassified points only elicit a slightly higher response in the wrong output neuron.



\subsection{Classifying multiple patterns}
\label{sec:class-mult-patt}
Here we show how it is possible to train a population of output neurons to classify multiple overlapping patterns in a supervised setting. For this demonstration we use the common benchmark of classifying handwritten digits from the MNIST data-set. Namely we test the system using MNIST digits from 0 to 4 scaled to $24 \times 24$ pixels, as in~\cite{Bill_Legenstein14}. In the network, there is an input layer consisting of $24 \cdot 24 \cdot n_\text{c}$ Poisson neurons, whose spike rates are scaled according to the intensity of the  MNIST digit image pixel, and the output layer consisting of 5 neurons that should learn to respond to the corresponding digit, and an additional layer of teacher neurons indicating which of 5 output neurons should fire in response to a given input. The index of the output neuron that fires the most in response to a test stimulus is considered the label that the network assigns to this input. During training 1000 digits, randomly drawn from the training set, are presented for $100$\,ms each while the learning circuits are enabled. The learning circuits are then disabled and the performance of the network is evaluated on 500 further digits (randomly drawn from the test set). Further implementation details are given in the supplementary material.

The learning algorithm is the same one used in Section~\ref{sec:single-patt-binary}. To compensate for the discretization errors, the update is made probabilistic as in~\cite{Muller_etal17}. Although we restrict ourselves to probabilistic signals that are independent per neuron, rather than per synapse, we achieve a performance comparable to that of~\cite{Bill_Legenstein14}.

\begin{figure}
\centering
	\begin{subfigure}{0.45\textwidth}
		\centering
		\includegraphics[width=\textwidth]{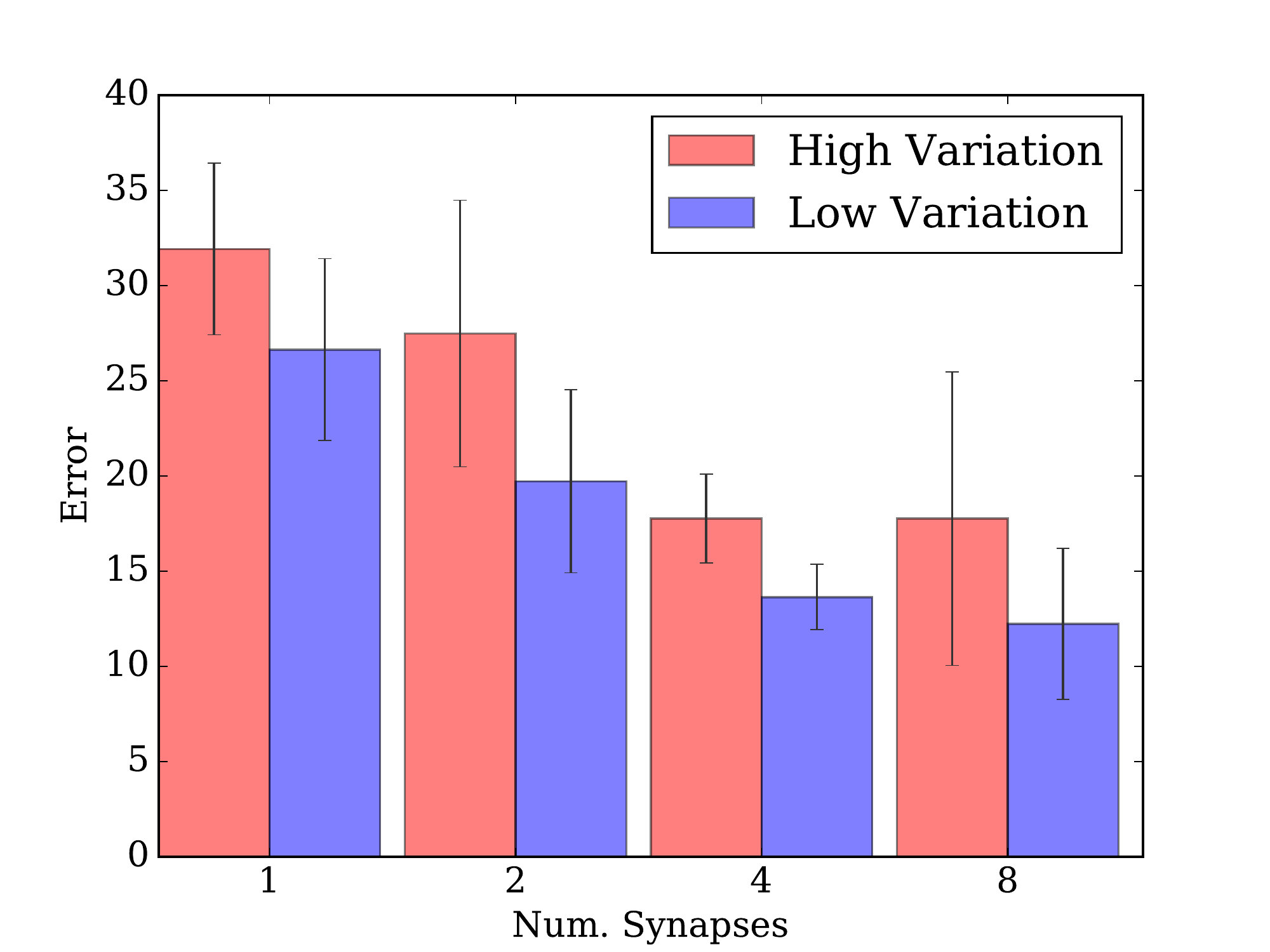}
		\subcaption[width=0.8\linewidth]{Error on the (reduced) MNIST classification task (test set) as a function of the number of binary synapses per input pixel for high and low CV. The CV settings correspond to Fig.~\ref{fig:mismatch}. Errorbars indicate std. dev. on 5 repetitions.} 
		\label{fig:performance}
	\end{subfigure}
	\quad
	\begin{subfigure}{0.45\textwidth}
		\centering
		\includegraphics[width=\textwidth]{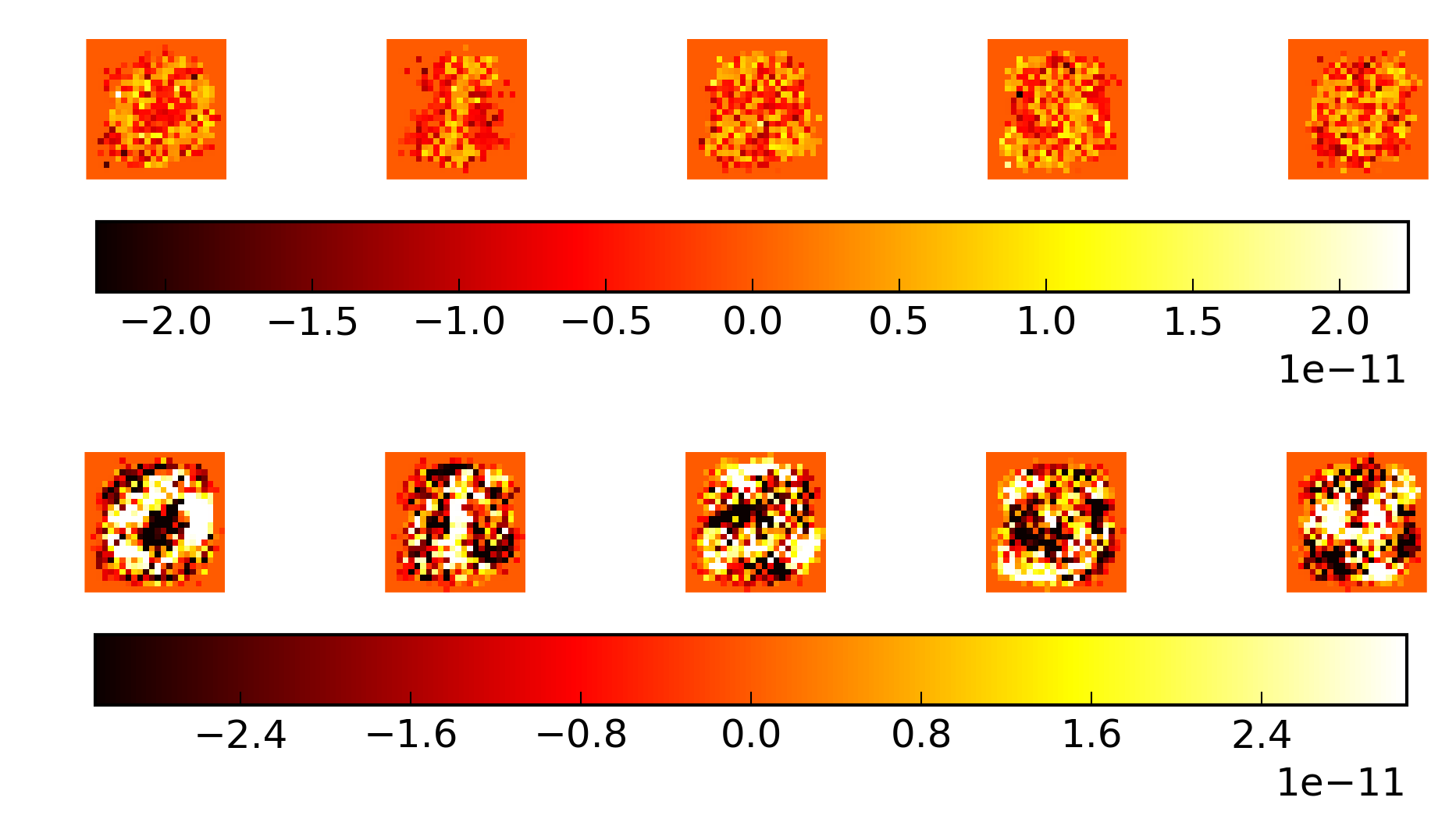}
		\subcaption[width=0.8\linewidth]{Learned weight matrices with one and eight synaptic weights.} 
		\label{fig:wplot}
	\end{subfigure}
\end{figure}

In Fig.~\ref{fig:performance} we report the performance of the network as a function of the number $n_c$ of synapses used per pixel,  in terms of classification accuracy. In Fig.~\ref{fig:wplot} we show two examples of the learned weight matrices. }{}

\revised{discussion}{
  \section{Discussion}
  \label{sec:discussion}

  The memristive synapse circuit proposed in this paper comprises two memristive devices, 20 \acp{MOSFET}s, and a pulse controller module.  Clearly, the resulting synaptic circuit is much larger in area than synapse elements employed in dense 1R or 1T-1R crossbar arrays~\cite{Garbin_etal15,Park_etal12b,Prezioso_etal15}. However, the large area overhead used allows the system to operate multiple synapses in parallel, both along the rows and along the columns of the synaptic array (e.g., by sending multiple \ac{AER} pulses in quick succession across multiple columns). In addition, as the currents passing through the memristive devices are contained in each individual synapse element and do not diffuse to neighboring devices, there are no sneak-path issues and no problems for quickly charging/discharging high capacitive loads. The strategy of using two memristive devices per synapse allows the use of a normalizer circuit, which has the desirable property of minimizing the effect of variability across the memristive devices. In addition, the strategy adopted to use the two devices in a differential way, increasing one while decreasing the other, eliminates the need for a precise reference for the normalizer operations, and provides automatically the possibility to implement both positive and negative synaptic weights in the network.
  
}{}

\section{Conclusion}
\label{sec:conclusion}
We proposed a differential current-mode memristive synapse circuit that decouples the current used to sense or change memristive device state from the current used to stimulate ultra low-power post-synaptic neuron circuits. \revised{concMN2}{We showed}{In addition, we showed} that the circuit  proposed  significantly reduces the effect of device variability, and that it is ideally suited for implementing advanced spike-based learning mechanisms that do not use overlapping pulses at the terminals of the target synapse. We argued that the strategy of using pulse extenders and Gilbert-normalizers in each synapse element maximizes throughput and minimizes power consumption in large-scale event-based neuromorphic computing platforms. For applications that do no require online adpatation or simultaneous read-write functionality, we describe how the synaptic circuit can be integrated with dense memristive crossbar arrays (Supplementary material). Given that memory-related constraints, such as size and throughput, represent one of the major bottlenecks in conventional computing architectures~\cite{Indiveri_Liu15}, and given the potential of neuromorphic computing platforms to perform robust computation using variable and slow computing elements, the proposed circuit offers an attractive solution for building alternative non von Neumann computing platforms with advanced and emerging memory technologies.


\section{Acknowledgements}
This work was supported by SNSF grant number CRSII2\_160756 and by the EU ICT grant ``NeuRAM$^3$'' (687299). 


\renewcommand{\bibfont}{\normalfont\small} 
\printbibliography 

\end{document}


\title[Supplementary Material]{Supplementary material:\\
  A differential memristive synapse circuit for on-line learning in neuromorphic computing systems}

\author{Manu V Nair, Lorenz K. Muller, and Giacomo Indiveri}

\address{Institute of Neuroinformatics, University of Zurich and ETH Zurich}
\ead{mnair$|$lorenz$|$giacomo@ini.uzh.ch}

\section{Learning Simulation Setup}
\label{sec:supplementary}

The following differential equations define our neuron model. They were derived in~\cite{Chicca_etal14} to describe the behavior of a neuron circuit that approximates an exponential adaptive integrate-and-fire neuron.
Compared to~\cite{Chicca_etal14} we added an additional term $I_\text{comp}$ that reflects the compensation current injected by the learning circuit, which was not present in the original equations. This current enforces that the neuron fires in accordance with its input teacher signal.
\begin{align}
\frac{dI_m(t)}{dt} &=
	\frac{I_\text{pos}(t) - I_m(t) \left( 1 + I_\text{adapt}(t) / I_\tau \right)}
    {\tau_m (1+ \frac{I_\text{th}}{I_m(t) + I_0} )} \\   
\frac{dI_\text{adapt}(t)}{dt} &=
	\frac{I_\text{p} - I_\text{adapt}(t) }{\tau_\text{adapt}} \\
I_m(t) &\leftarrow I_\text{reset} \text{   if } I_m(t) > I_\text{spkthr}
\end{align}

With synaptic differential equation
\begin{align}
\frac{I_\text{syn}(t)}{dt} = -I_\text{syn}(t) + I_w w_\text{syn} \sum_i \delta(t^\text{spike}_i - t).
\end{align}

For compactness we introduced
\begin{align}
I_\text{pos}(t) &=
	I_\text{fb}(t) + \frac{I_\text{th}}{I_\tau} 
    \left(I_\text{in}(t) + I_\text{comp}(t) + I_\text{Syn}(t) - I_\text{adapt}(t) - I_\tau \right) \\
I_\text{fb}(t)  &=
	\frac{I_a(t)}{I_\tau}  (I_m(t) + I_\text{th}) \\
I_a(t)   &=
	\frac{I_g}{ 1 + \exp \left(-(I_m(t)-I_\text{ath})/I_\text{anorm}\right) } 
\end{align}

The table \ref{tab:neu} clarifies the meanings and where appropriate the values of the above variables and parameters.

\begin{table}
\centering
\begin{tabular}{ l | l | l }
  Name & Description & Value \\
  \hline			
  $I_m(t)$ & Membrane Current & Variable \\
  $I_\text{pos}(t)$ & Positive input to the neuron & Variable \\
  $I_\text{adapt}(t)$ & Adaptation feedback & Variable \\
  $I_\tau$ & Neuron time constant bias & 2 pA \\
  $\tau_m$ & Neuron time constant & 8.9 ms \\
  $I_\text{th}$ & Global synaptic input scaling factor & 1 pA \\
  $I_0$ & Leak Current (process parameter) & 0.5 pA \\
  $I_\text{p}$ & Feedback rate & 0.5 pA \\
  $\tau_\text{adapt}$ & Adaptation time constant & 17.7 ms\\
  $I_\text{reset}$ & Membrane reset level & 1 pA\\
  $I_\text{spkthr}$ & Spike threshold & 60 pA\\
  $I_\text{syn}$ & Synaptic input current & Variable\\
  $n_{in}$ & no. synapses per input channel & given in plots\\
  $I_w$ & Synaptic bias current & 16 pA (MNIST)\\
  $I_w$ & Synaptic bias current & 1 nA $/ n_{in}$ (single pattern)\\
  $w_\text{syn}$ & Synaptic Weight & Drawn from distributions \\ 
  & & given for device variability \\
  $t^\text{spike}_i$ & Times of presynaptic spikes & Variable\\
  $I_\text{fb}(t)$ & Feedback current & Variable \\
  $I_a (t)$ & Fitting variable for feedback & Variable \\
  $I_g$ & Feedback gain & 1 nA \\
  $I_\text{ath}$ & Fitting variable & 20 nA\\
  $I_\text{anorm}$ & Fitting variable & 1 nA\\
  \hline  
\end{tabular}
\caption{Neuron and Synapse Variables and parameters.}
\label{tab:neu}
\end{table}

\subsection{Learning Equations}
The learning block acts according to:
\begin{align}
\frac{dT(t)}{dt} &=  \frac{-T(t)}{\tau_\text{learn}} \\
\frac{dS(t)}{dt} &=  \frac{-S(t)}{\tau_\text{learn}} \\
I_\text{comp}(t) &=  g_\text{comp} (T(t) - S(t)) ( t < t_\text{stop})\\ 
q(t) &= (S(t) + S_0 - I_\text{syn}(t)) \\
L(t) &= \text{sign}( q(t) (|q(t)| > \alpha) ) b_p(t)
\end{align}

When a presynaptic spike arrives, $w_\text{syn}$ is updated according to the value of $L(t)$. The update is to redraw $w_\text{syn}$ from one of the two distributions given in the main paper for high and low coefficient of variation, rescaled so that both have the same mean. For the single pattern simulations the distributions were normal distributions with the same mean and variance as the aformentioned. Overall this results in a probabilistic binary update.

In the case of the high resolution synapse the last equation becomes an analog update proportional to $q(t)$ (with scaling factor 0.0001).

On a postsynaptic spike / a teacher spike $w_\text{S}$ / $w_\text{T}$ are instantaneously added to $S(t)$ / $T(t)$.

This implements an update that is similar to \cite{Urbanczik_Senn14}; the key differences are the use of a sign function to limit the output of the circuit to $\lbrace-1,0,1\rbrace$ (decrease, keep, increase synaptic value) and the addition of $b_p(t)$ that probabilistically enables / disables the learning block to implement a form of RUSD \cite{Muller_etal17}. 

The functional effect of these equations is to perform a gradient descent procedure of brining the output spike rate of the neuron close to the teacher signal (as shown in \cite{Urbanczik_Senn14}), but in contrast to \cite{Urbanczik_Senn14} updates are discretized and probabilistic as in RUSD \cite{Muller_etal17}.

The table \ref{tab:learn} clarifies the meanings and where appropriate the values of the above variables and parameters.

\begin{table}
\centering
\begin{tabular}{ l | l | l }
  Name & Description & Value \\
  \hline			
  $T(t)$ & Teacher signal & Variable \\
  $\tau_\text{learn}$ & learning time constant & 8 ms \\
  $S(t)$ & Instantaneous firing rate & Variable\\
  $I_\text{comp}$ & Compensation current from learning block & Variable \\
  $g_\text{comp}$ & Compensation gain & 1.0 \\
  $t_\text{stop}$ & Time at which teaching input ends & $n_\text{samp} \cdot \tau_\text{samp}$ \\
  $q(t)$ & Cost function & Variable\\
  $S_0$ & Output bias & $(-500, 0)$ pA (MNIST, single pattern) \\ 
  $L(t)$ & Learning circuit output & Variable \\
  $\alpha$ & Slack variable & $(300, 500)$ pA (MNIST, single pattern) \\
  $b_p(t)$ & Bernoulli learning variable & $p=(0.01,0.001)$ (MNIST, single pattern) \\
  $w_\text{S}$ & Output spike weight & 200 pA \\
  $w_\text{T}$ & Teacher spike weight & 40 pA\\
  \hline  
\end{tabular}
\caption{Learning circuit variables and parameters.}
\label{tab:learn}
\end{table}

\subsection{Population Level Parameters}
The table Tab. \ref{tab:pop} gives an overview of the values population level parameters were set to. 

\begin{table}
\centering
\begin{tabular}{ l | l }
  Description & Value \\
  \hline			
  Firing rate positive single pattern & $(x\cdot$50 kHz$ + $5 kHz$)/n_{in}$; $x \sim U(0.5,1)$\\
  Firing rate negative single pattern & $(x\cdot$50 kHz$ + $5 kHz$)/n_{in}$; $x \sim U(0,0.5)$\\
  Teacher rate positive single pattern & 50 kHz\\
  Teacher rate negative single pattern & 0 kHz\\
  Firing rate MNIST &  $z\cdot$100 Hz; $z$ is the pixel value \\
  Teacher rate positive MNIST class & 100 Hz\\
  Teacher rate negative MNIST class & 0 Hz\\
  No. teacher neurons & 40 \\
  \hline  
\end{tabular}
\caption{Population level parameters.}
\label{tab:pop}
\end{table}

\section{Advantages of the normalizer circuit over a current mirror}
\label{sec:currmirror}
\begin{figure}[ht]
 	\centering
 	\includegraphics[width=0.4\textwidth]{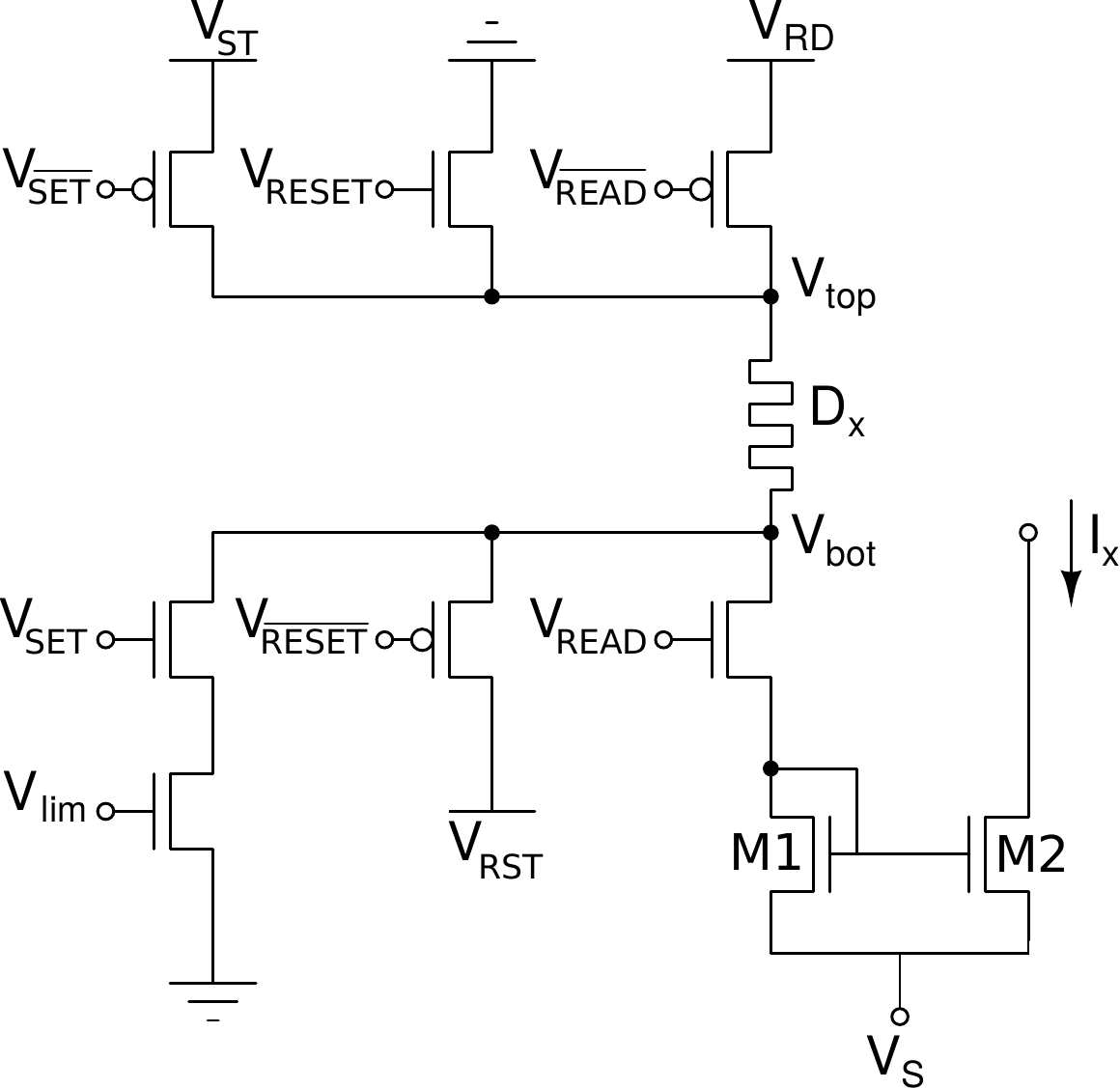}
 	\caption{A current mirror synaptic memory cell with read and write functionality}
 	\label{fig:currmirror}
 \end{figure}

 Isolating the current flowing through the device from that sent into the post-synaptic neuron can also be achieved by use of a simple current mirror circuit as shown in Fig.~\ref{fig:currmirror}. This circuit is put in read mode when $Read$ and $\overline{Read}$ are set to logical 1 and 0 values, respectively.
 However, with this circuit the $\frac{Width}{Length}$ ratio between the transistors M1 and M2 will have to be made very large to reduce the output current $I_x$. For example, to reduce the read current from $1\mu A$ to $1 nA$, which is a typical value for input currents to sub-threshold silicon neurons~\cite{Indiveri_etal11}, would require a ratio of 1000. Secondly, without use of power-hungry active feedback circuit, the drain voltages of M1 and M2 will be different, which will result in poor current mirroring accuracy. The normalizer circuit we propose addresses both of these issues. The proposed circuit is also more compact. This is because while the normalizer uses two extra transistors (M5 and S13 in Fig.~\ref{fig:multi_mem}), these are the sized comparably to M1-M4. The absence of large scaling factors makes the total area of the circuit smaller. Note that while the differential synapse circuit comprises two copies of the basic current mirror, it also provides twice the dynamic range. Finally, the proposed design can be extended in various ways as described in the following sections.
 
 \section{Improving linearity}
 \label{sec:linearity}
\begin{figure}[ht]
 	\centering
 	\includegraphics[width=0.8\textwidth]{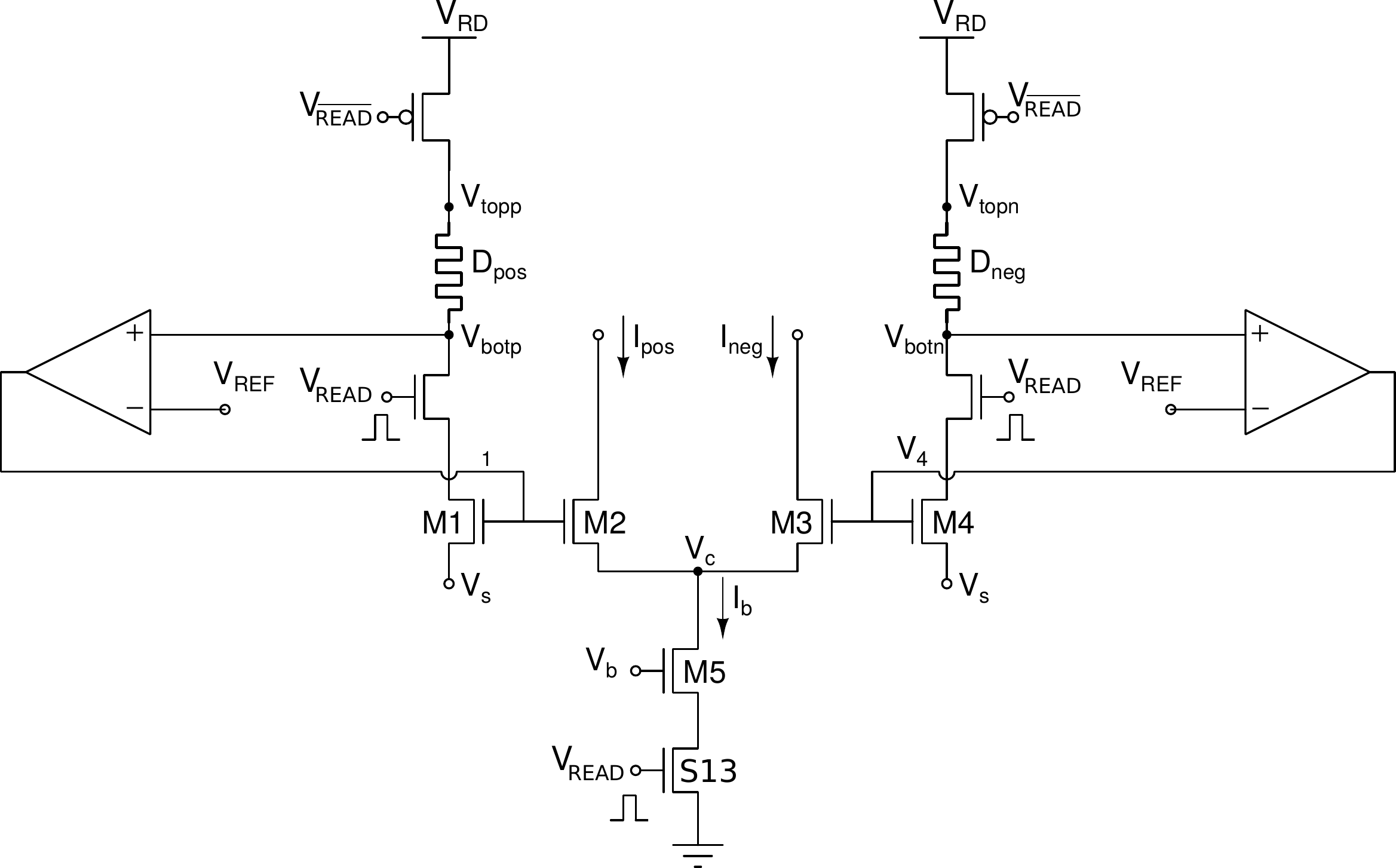}
 	\caption{Active circuit schematics to improve linearity. Only the read path is shown. The write circuit remains the same as illustrated in Fig~\ref{fig:switches}.}
 	\label{fig:diff_mem_act}
 \end{figure}

 Equation~\ref{eq:inputcurrent} is obtained by making an approximation to the transcendental Eq.~\ref{eq:dioderes}. The approximation error can be eliminated by use of active elements as shown in Fig.~\ref{fig:diff_mem_act}.
 By use of feedback circuits, the voltage at the bottom nodes of the memristive devices are set to $V_{REF}$. Therefore, the current flowing through M1 and M4 is made a linear function of the device conductance. The normalizer outputs are consequently a linear function of the device conductance.
 \begin{align}
 I_{M1} &= G_{pos} \cdot (V_{RD} - V_{REF}) \\
 I_{M4} &= G_{neg} \cdot (V_{RD} - V_{REF}) \\
 I_{pos} &= I_b \cdot \frac{I_{M1}}{I_{M1}+I_{M4}} = I_b \cdot \frac{G_{pos}}{G_{pos} + G_{neg}}\\
 I_{neg} &= I_b \cdot \frac{I_{M4}}{I_{M1}+I_{M4}} = I_b \cdot \frac{G_{neg}}{G_{pos} + G_{neg}}
 \end{align}
 While the active circuits will consume additional power and area, it improves linearity. Additionally, by making the difference $V_{RD} - V_{REF}$ small, the currents $I_{M1}$ and $I_{M4}$ can be made much smaller, partly compensating for the increased power consumption resulting from the use of Op-Amps.
 
 \section{Multi-device memory block}
 \label{sec:multidev}
\begin{figure}[ht]
 	\centering
 	\includegraphics[width=\textwidth]{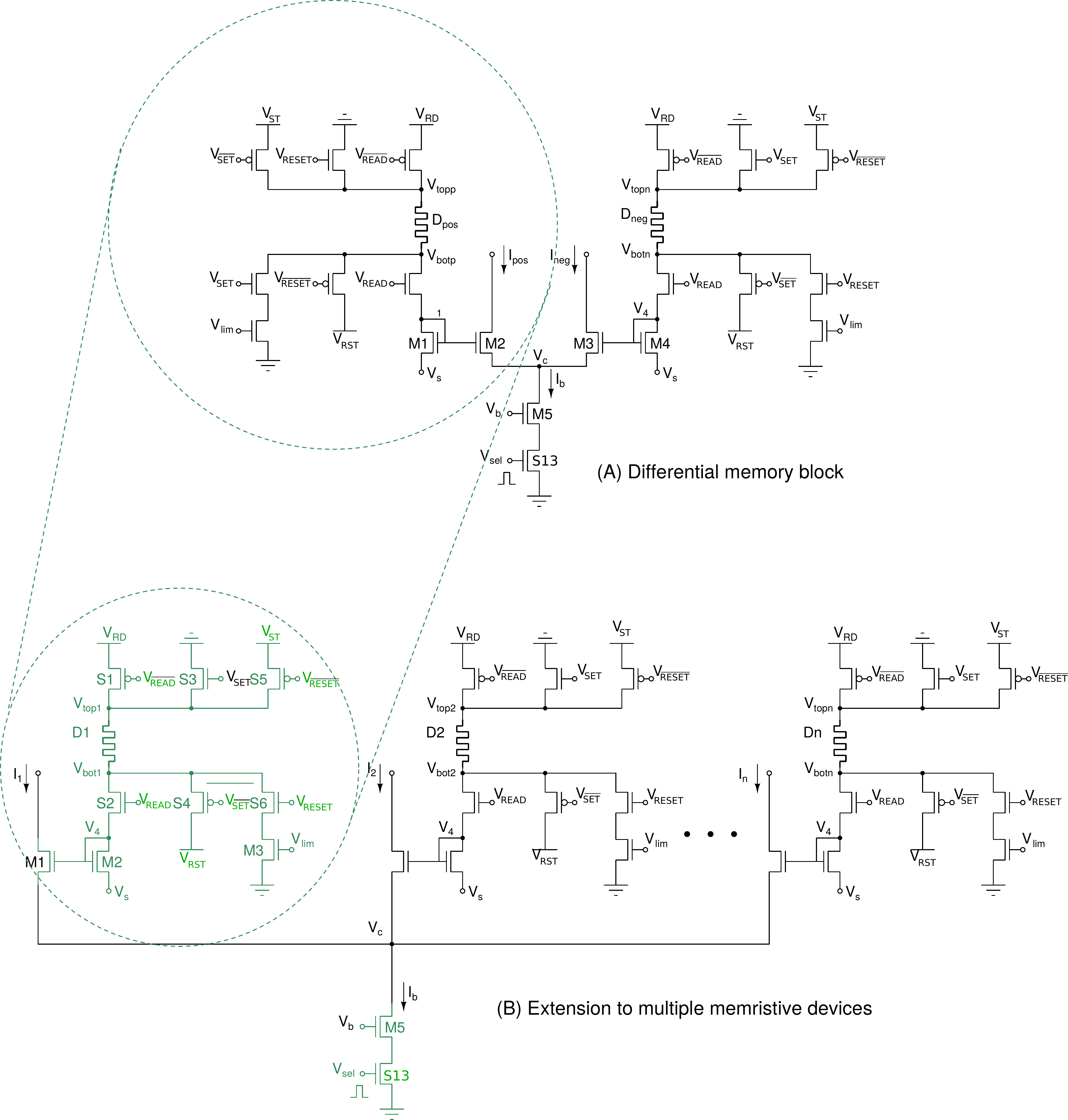}
 	\caption{Extension to multiple memory devices}
 	\label{fig:multi_mem}
 \end{figure}

 The proposed normalizer circuit can be extended to multiple memristive devices. This can be used to create multi-state memory cells using devices that have only two stable states. Figure~\ref{fig:multi_mem} illustrates this for $n$ devices where each branch comprises the arrangement highlighted with green circles. The currents generated by each branch are approximately equal to:
 \begin{align}
 \label{eq:multi}
 I_k &\approx I_b \cdot \frac{G_k}{\Sigma_{x=1}^{n} G_x}
 \end{align}
 where, $G_k$ is the conductance of device $D_k$.
 
 These output currents can be combined to create positive and negative output currents in different ways, depending on the application. For example, 
 \begin{align}
 \label{eq:simple}
 I_{pos} &= \sum\limits_{x=1}^{m-1} I_x \\
 I_{neg} &= \sum\limits_{x=m}^{n} I_x
 \end{align}
 
 Another possibility to expand the dynamic range of the output currents is to weigh the currents output from each branch as follows:
 \begin{align}
 I_{pos} &= \sum\limits_{x=1}^{m-1} 2^x \cdot I_x \\
 I_{neg} &= \sum\limits_{x=m}^{n} 2^{x-m} \cdot I_x
 \end{align}
 
 \subsubsection*{Read operation}: The read operation of multi-memory cell is enabled when the $V_{sel}$, $Read$, and $\overline{Read}$ signals are set to logical 1, 1 and 0, respectively. In this mode, transistors S1 and S2 in each branch shown in Fig.~\ref{fig:multi_mem} are on, and transistors S3-S6 are off. In this configuration, each branch in the circuit shown in Fig.~\ref{fig:multi_mem} generates a current given by Eq~\ref{eq:multi}.
 
 \subsubsection*{Write operation}: As in the case of the two device differential cell, programming the state of the devices is achieved by suitably setting the corresponding $Setx$ and $Resetx$ signals of the branch. Increasing or decreasing the synaptic weight is achieved by modifying the conductances of a subset of devices contributing to $I_{pos}$ and $I_{neg}$. The specific control circuitry for determining $Setx$ and $Resetx$ signals depends on the specific equations used to generate the output currents, $I_{pos}$ and $I_{neg}$.
 
 \section{Integration into a crossbar array}
 \label{sec:xbar}
There are several applications where area is an important concern, and dense 2D synaptic arrays are desired. 
 
 \begin{figure}[ht]
 	\centering
 	\includegraphics[width=\textwidth]{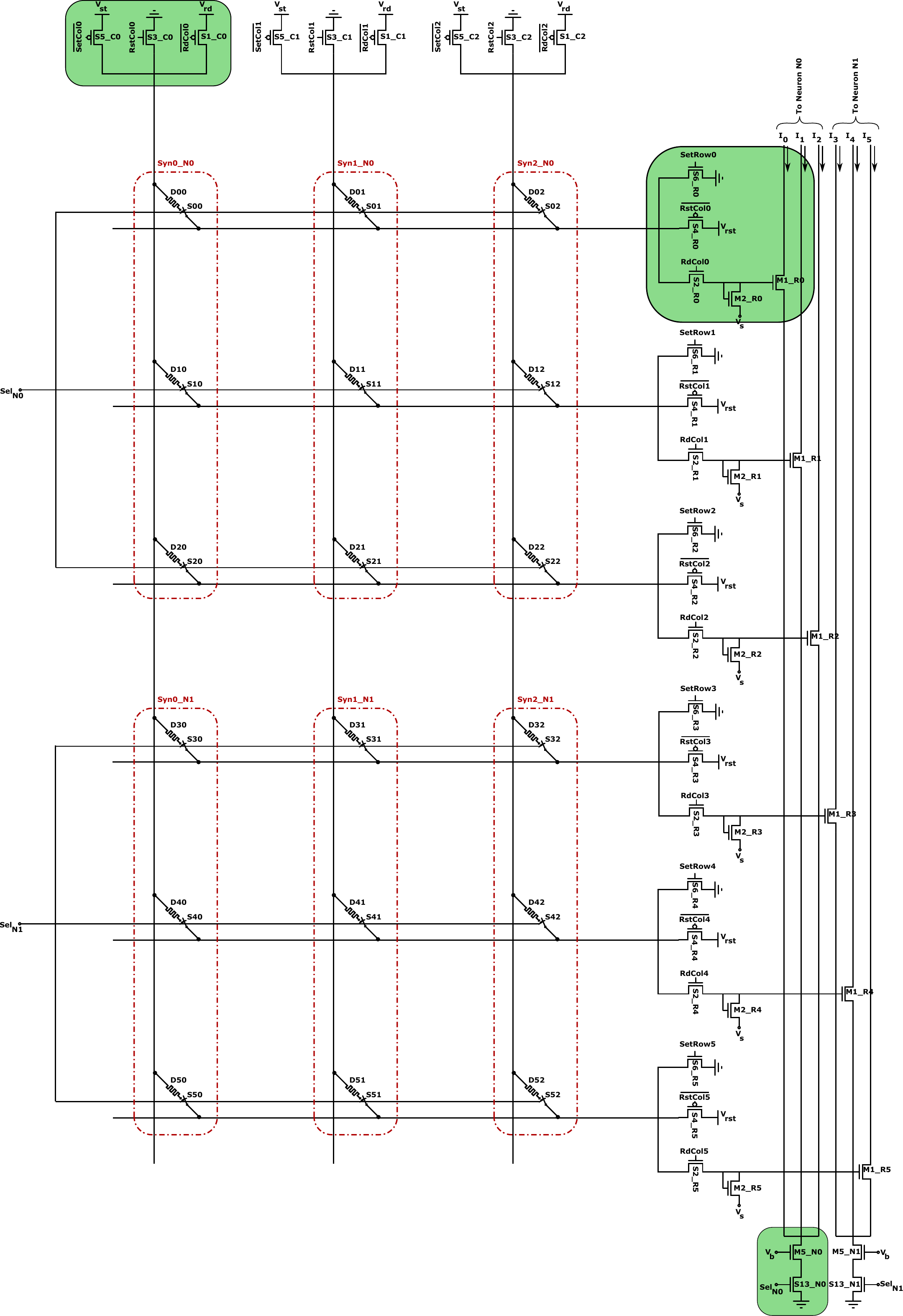}
 	\caption{The differential memory circuit in a crossbar array. The switches $Sxy$ can be eliminated to increase crossbar density without affecting the functionality of the circuit at the cost of increased power consumption.}
 	\label{fig:xbar}
 \end{figure}
 
 The circuits presented can be integrated in a dense crossbar array by taking apart the circuit shown in Fig~\ref{fig:multi_mem} and rearranging them as illustrated in Fig~\ref{fig:xbar}. The crossbar circuit shown in Fig~\ref{fig:multi_mem} has 6 synapses implemented in a $6 \times 3$ crossbar array connected to two neurons, where each synapse comprises a three device memory cell. The circuit elements implementing the proposed normalizer circuits are highlighted in green and labeled to match to the corresponding circuit elements in Fig.~\ref{fig:multi_mem}. In this design, circuits elements are shared by multiple synapses. That is, all the memristive devices in a column, Cx, share the transistors S1\_Cx, S3\_Cx, and S5\_Cx. All the memristive devices in a row, Rx, share the transistors S2\_Rx, S4\_Rx, and S6\_Rx. The normalizer bias circuits comprising, M5\_Nx and S13\_Nx, are shared by all synapses associated with neuron Nx. When a synapse, Synx\_Ny is read, the three devices in it are put in a configuration equivalent to the one shown in Fig.~\ref{fig:multi_mem} with $n=3$. In Fig~\ref{fig:xbar}, each memristive device, Dxy, has a series access transistor, Sxy, that prevents flow of current through inactive synapses. This transistor can be eliminated to increase crossbar density without affecting the functionality of the circuit at the cost of increased power consumption.
 
 \subsubsection*{Read operation}: When a spike train addressed to neuron $Ni$ arrives, the corresponding $Sel_{Ni}$ signal turns on the the switches Sxy of the synapses afferent to the neuron. Transistors S1\_Cx and S2\_Ry corresponding to the neuron $Ni$ are also turned on. This connects all the synapses belong to neuron $Ni$ to its normalizer bias circuit. By the principle of linear super-position. Even though all synapses share the same normalizer bias block, the current read out of the normalizer is the same as that generated if each synapse had its own normalizer block.
 
 \subsubsection*{Write operation}: At the end of the read operation, all the synapses are updated as per the directions of the learning block or the downstream neuron. However, unlike the differential memristive synapse circuit, described in Sec~\ref{sec:diff-memr-synapse}, a subset of the synapses cannot be programmed while part of it is being read. The write operation in this arrangement requires a controller that issues a sequence of signals that gates all read activity while the state of the crossbar array is being updated. The entire array can be updated in two phases. In the first phase, all those device whose conductances need to be increased are updated by enabling to corresponding $SetColx$ and $SetRowy$ signals. In the next phase, the devices whose conductances are to be decreased are programmed by programming the $ResetColx$ and $ResetRowy$ signals. Similar update schemes have also been proposed in earlier works such as~\cite{Nair_Dudek15,Prezioso_etal15}.

\renewcommand{\bibfont}{\normalfont\small} 
\printbibliography